\pdfoutput=1

\documentclass[a4paper,twoside]{article}

\newcommand{\affil}[1]{$^{\rm #1}$}

\usepackage{graphicx,amssymb}
\usepackage{epsf,psfig}
\usepackage{txfonts}
\usepackage{multirow}
\usepackage{bigstrut}
\usepackage{bigdelim}
\usepackage{placeins}
\usepackage{amstext}

\date{\small Received: 9, March 2012; Accepted: 24, August 2012; Online publication: 20, February 2013}

\title{\large\bf\flushleft Revealing the character of orbits in a binary system consisting of a primary
galaxy and a satellite companion}

\author{\parbox{\textwidth}{\flushleft
\vspace{-0.5 cm}
{\it Euaggelos E. Zotos\affil{}}\\
\vspace{0.4 cm}
{\small \affil{}\,Department of Physics,
Section of Astrophysics, Astronomy and Mechanics,
Aristotle University of Thessaloniki, \\
GR-541 24, Thessaloniki, Greece}\\
{\small \affil{}\,Email: evzotos@physics.auth.gr}}}

\begin{document}

\twocolumn[
\begin{changemargin}{.8cm}{.5cm}
\begin{minipage}{.9\textwidth}
\vspace{-1cm}
\maketitle

\small{\bf Abstract:}

In this article, we present a galactic gravitational model of three degrees of freedom (3D), in order to study and reveal the character of the orbits of the stars, in a binary stellar system composed of a primary quiet or active galaxy and a small satellite companion galaxy. Our main dynamical analysis will be focused on the behavior of the primary galaxy. We investigate in detail the regular or chaotic nature of motion, in two different cases: (i) the time-independent model in both 2D and 3D dynamical systems and (ii) the time-evolving 3D model. For the description of the structure of the 2D system, we use the classical method of the Poincar\'{e} $(x,p_x), y=0, p_y>0$ phase plane. In order to study the structure of the phase space of the 3D system, we take sections in the plane $y=0$ of the 3D orbits, whose initial conditions differ from the plane parent periodic orbits, only by the $z$ component. The set of the four-dimensional points in the $\left(x, p_x, z, p_z\right)$ phase space is projected on the $(z, p_z)$ plane. The maximum Lyapunov characteristic exponent is used in order to make an estimation of the chaoticity of our galactic system, in both 2D and 3D dynamical models. Our numerical calculations indicate that the percentage of the chaotic orbits increases when the primary galaxy has a dense and massive nucleus. The presence of the dense galactic core also increases the stellar velocities near the center of the galaxy. Moreover, for small values of the distance $R$ between the two bodies, low-energy stars display chaotic motion, near the central region of the galaxy, while for larger values of the distance $R$, the motion in active galaxies is entirely regular for low-energy stars. Our simulations suggest that in galaxies with a satellite companion, the chaotic nature of motion is not only a result of the galactic interaction between the primary galaxy and its companion, but also a result caused by the presence of the dense nucleus in the core of the primary galaxy. Theoretical arguments are presented in order to support and interpret the numerically derived outcomes. Furthermore, we follow the 3D evolution of the primary galaxy, when mass is transported adiabatically from the disk to the nucleus. Our numerical results are in satisfactory agreement with observational data obtained from the M51-type binary stellar systems. A comparison between the present research and similar and earlier work is also made.

\medskip{\bf Keywords:}

Galaxies: kinematics and dynamics; interacting galaxies

\medskip
\medskip
\end{minipage}
\end{changemargin}
]
\small

\section{Introduction}

It is well known that galaxies are not observed as isolated entities, but often appear in binary, triple, or multiple stellar systems. On the contrary, just as there are clusters of stars, there are also clusters of galaxies. This leads to gravitational galactic interactions. The Milky Way--Magellanic Clouds system is a very indicative example of galactic interaction (see Gardiner, Sawa, \& Fujimoto 1994; Lin, Jones, \& Klemola 1995; Weinberg 1995, 1998; Van der Marel 2001; Putman et al. 2003; Connors et al. 2004; Bekki \& Chiba 2005). Another interesting stellar system of interacting galaxies is the Andromeda galaxy M31, with its two smaller companion galaxies M32 and NGC 205. The gravitational effects of the M32 galaxy on the spiral structure of M31 were studied by Byrd (1978). This interesting triple stellar system was also investigated in an earlier work, using a self-consistent computer simulation code with revealing results (see Vozikis \& Caranicolas 1994).

One of the main reasons for the transition from ordered to chaotic motion is the presence of massive objects in the central regions of the galaxies. Stars reaching the galactic center on highly eccentric redial orbits can be scattered out of the galactic plane displaying chaotic motion (see Sellwood \& Moore 1999). Furthermore, a central mass concentration can strongly perturb the stellar box orbits in elliptical galaxies which become chaotic (Merritt 1996). Observational indications suggest the presence of a strong central mass concentration by a very sharply rising rotation curve. A second reason for chaotic behavior in the galactic motion is the presence of strong external perturbations (see Zotos 2011c). Increasing the magnitude of perturbations weakens or destroys the stability of orbits, thereby increasing the percentage of the stochasticity in the stellar system. Resonances are also responsible for chaotic motion (see Contopoulos \& Grosb{\o}l 1986; Cincotta, Giordano, \& Perez 2006). Of course, resonances are responsible not only for the presence of chaos in galactic stellar systems but also for the chaotic motion in the solar system (see Wisdom 1987; Henrard \& Caranicolas 1990). The reader can find interesting information about the chaotic motion in galaxies and its connection with observational data in Grosb{\o}l (2002).

Therefore, it seems very challenging to construct a galactic gravitational model of three degrees of freedom (3D), in order to study and reveal the character of the orbits of the stars, in a binary stellar system composed of a primary quiet or active galaxy and a small satellite companion galaxy. The aim of the present paper is to study the character (regular or chaotic) of motion in the primary galaxy (hereafter the galaxy) under the perturbation caused by a satellite galaxy and to connect the degree of chaos with parameters such as the mass of the nucleus of the galaxy or the distance between the two galaxies. We shall also study the behavior of the velocities near the central region and try to connect this behavior with the scale length of the nucleus. Furthermore, we shall deal with the case where adiabatic mass transfer takes place in the galaxy, that is, the galaxy evolves with time. In this initial research, only the case where the two bodies move in circular orbits about the center of mass of the system is considered, for simplicity. We plan to study the case of the satellite in inclined, elliptic orbits in a future work.

In a recent paper (Zotos 2012b), we investigated in detail and revealed the regular or chaotic character of motion in a galactic gravitational model of three degrees of freedom describing a binary quasar system. According to the model, two quasars are hosted in a pair of interacting disk galaxies. Here we must emphasize that in the present research the second body of the binary stellar system, that is, the satellite companion galaxy, is considered and treated as a point mass. On the contrary, in Zotos (2012b), we deal with a much more complicated binary system, since none of the two interacting disk galaxies can be treated as point mass due to the fact that both of them are hosting quasars in their cores. Thus, we may say that Zotos (2012b) is a reasonable extension and generalization of the present paper. Taking into account that in both papers we study the properties of motion of the stars in similar, but in no case identical binary systems, it is natural and well expected to use the same philosophy regarding the basic setup of the gravitational models, the handling of the numerical calculations, and also the same standard methods of galactic dynamics and celestial mechanics.

The present article is organized as follows. In Section 2, we present our gravitational dynamical model, which describes the motion in a binary stellar system. In Section 3, we provide an analysis of the two degrees of freedom (2D) system, considering orbits in the galactic plane $(z=0)$. In Section 4, we study the character of motion in the 3D system, using different kinds of dynamical methods. In Section 5, we present an interesting theoretical analysis in order to support the numerically obtained results. In Section 6, we use a 3D time-dependent model in order to follow the evolution of orbits as mass is transported adiabatically from the disk to the nucleus of the primary galaxy. We close this paper with Section 7, where the conclusion is presented and a comparison between the present theoretical results with observational data is made. In the same section, we include statements concerning the astrophysical relevance of this research work.

\section{Presentation of the Dynamical Model}

Our gravitational model describes the dynamical properties of a binary stellar system which is composed of a primary galaxy and a small satellite companion galaxy. The primary galaxy (hereafter the galaxy) consists of a disk and a spherical nucleus and is described by the potential
\begin{eqnarray}
V_{\rm P}(r,z) &=& V_{\rm d}(r,z) + V_{\rm n}(r,z) \nonumber \\
&=& - \frac{G M_{\rm d}}{\sqrt{b^2 + r^2\!+\! \left(a + \sqrt{h^2 + z^2}\right)^2}} \nonumber \\
&-& \frac{G M_{\rm n}}{\sqrt{r^2 + z^2 + c_{\rm n}^2}}\,,
\end{eqnarray}
where $r^2=x^2+y^2$, $M_{\rm d}$ and $M_{\rm n}$ is the mass of the disk and the nucleus of the primary galaxy respectively, $a$ is the disk's scale length, $h$ is the disk's scale height, $b$ is the core radius of the disk halo, while $c_{\rm n}$ is the scale length of the nucleus. The disk of the primary galaxy is represented by the well-known Miyamoto--Nagai model (Miyamoto \& Nagai 1975), with an additional core radius parameter (see Carlberg \& Innanen 1987). The additional core radius is used so that the rotation curve displays high values at large distances from the galactic center, justifying in a way the presence of dark matter. The Plummer sphere we choose to describe the nucleus has been used many times in the past, in order to study the effects of the introduction of a central mass component in the core of a galaxy (see Hasan \& Norman 1990; Hasan et al. 1993). The satellite galaxy is described by a point-mass potential
\begin{equation}
V_{\rm S}(r,z) = - \frac{G M_{\rm s}}{d},
\end{equation}
where $d^2=x^2+y^2+z^2$, while $M_{\rm s}$ is the mass of the satellite galaxy. Therefore, the total gravitational potential describing the motion in our binary system is $V_{\rm T}(r,z) = V_{\rm P}(r,z) + V_{\rm S}(r,z)$.

In our study, we shall use the theory of the circular restricted three-body problem. The two bodies move in circular orbits in an inertial frame \(OXYZ\), with the origin at the center of mass of the system, with a constant angular frequency $\Omega_p > 0$, given by Kepler's third law
\begin{equation}
\Omega_p= \sqrt{\frac{G M_{\rm t}}{R^3}},
\end{equation}
where $M_{\rm t} = M_{\rm d} + M_{\rm n} + M_{\rm s}$ is the total mass of the binary system, while $R$ is the distance between the centers of the two galaxies. A clockwise, rotating frame \(Oxyz\), is used with the axis \(Oz\) coinciding with the axis \(OZ\) and the axis \(Ox\) coinciding with the straight line joining the two bodies. In this frame, which rotates with angular frequency $\Omega_p$, the two galactic centers have fixed positions $C_1(x,y,z) = \left(x_1,0,0\right)$ and $C_2(x,y,z) = \left(x_2,0,0\right),$ respectively. The total potential which is responsible for the motion of a star in the dynamical system of this binary system is
\begin{equation}
\Phi_{\rm t}(x,y,z) = \Phi_{\rm P}(x,y,z) + \Phi_{\rm S}(x,y,z) + \Phi_{\rm rot}(x,y),
\end{equation}
where
\begin{eqnarray}
\Phi_{\rm P}(x,y,z) &=& - \frac{G M_{\rm d}}{\sqrt{b^2 + r_{a1}^2 + \left(a + \sqrt{h^2 + z^2}\right)^2}} \nonumber \\
&-& \frac{G M_{\rm n}}{\sqrt{r_1^2 + c_{\rm n}^2}}, \nonumber \\
\Phi_{\rm S}(x,y,z) &=& - \frac{G M_{\rm s}}{r_2}, \nonumber \\
\Phi_{\rm rot}(x,y) &=& -\frac{\Omega_p^2}{2}\left[\frac{M_{\rm s}}{M_{\rm t}}r_{a2}^2 + R_{\rm s} r_{a1}^2 -
R^2 \frac{M_{\rm s}}{M_{\rm t}}R_{\rm s} \right],
\end{eqnarray}
and
\begin{eqnarray}
r_{a1}^2 &=& \left(x-x_1\right)^2 + y^2, \nonumber \\
r_{a2}^2 &=& \left(x-x_2\right)^2 + y^2, \nonumber \\
r_1^2 &=& r_{a1}^2 + z^2, \nonumber \\
r_2^2 &=&  r_{a2}^2 + z^2,
\end{eqnarray}
with
\begin{eqnarray}
x_1 &=& -\frac{M_{\rm s}}{M_{\rm t}}R, \nonumber \\
x_2 &=& R \left(1 - \frac{M_{\rm s}}{M_{\rm t}}\right) = R + x_1, \nonumber \\
R_{\rm s} &=& 1 - \frac{M_{\rm s}}{M_{\rm t}}.
\end{eqnarray}

The angular frequency $\Omega_p$ is calculated as follows. The two bodies circulate around their common mass center of the system with angular frequencies $\Omega_{p1}$ and $\Omega_{p2}$ given by
\begin{eqnarray}
\Omega_{p1} &=& \sqrt{\frac{1}{x_1}\left(\frac{-\text{d} V_{\rm S}(r)}{\text{d}r}\right)_{r=R}}, \nonumber \\
\Omega_{p2} &=& \sqrt{\frac{1}{x_2}\left(\frac{\text{d} V_{\rm P}(r)}{\text{d}r}\right)_{r=R}}.
\end{eqnarray}
As the primary galaxy cannot be considered as a mass point, the two angular frequencies are not equal, in general. However, this issue can easily be resolved. The angular frequencies of the two bodies can become equal under reasonable assumptions. This is justified if the final set of the parameters has physical meaning and represents satisfactorily the dynamical system. The equation $\Omega_{p1}=\Omega_{p2}$ leads to a fourfold infinity of solutions in the four unknowns $\left(a, b, h, c_{\rm n} \right)$. If one chooses proper values (representing the dynamical system), say, for the three parameters $\left(b, h, c_{\rm n} \right)$, then the equation $\Omega_{p1}=\Omega_{p2}$ gives only two values for the parameter $a$. One value is positive, while the other is negative and is rejected. The author would like to make it clear that after choosing properly the parameters, the deviation between the two angular frequencies is negligible, so that $\nu =
\left(| \Omega_{p1}-\Omega_{p2} |\right)/\Omega_{p1}$ is of the order of $10^{-8}$ or even smaller and $\xi = | \Omega_{p1}-\Omega_p |$ or $ \xi = | \Omega_{p2}-\Omega_p |$ is of the order of $10^{-6}$. Therefore, we consider the two angular frequencies almost equal, that is $\Omega_{p1}=\Omega_{p2}=\Omega_p$. Moreover, the treatment of large bodies with spherical symmetry as mass points is very common in celestial modeling. This method is quite familiar to those measuring the masses of disk galaxies.

The above setup in the rotating frame has been applied successfully in several previous papers studying the nature of binary stellar systems (see Caranicolas \& Papadopoulos 2009; Caranicolas \& Zotos 2009a; Caranicolas \& Innanen 2009; Zotos 2012b). Thus, we may conclude that the dynamical model described in Equations (4) and (5) is a good and also realistic model for a pair of interacting galaxies at the given mass ratio. Furthermore, in a recent paper (Zotos 2012b), we used the same treatment and setup, in order to study the dynamical properties in a binary system of two interacting galaxies. This could be considered a reasonable extension of the present study, where the second galaxy here is assumed to be only a satellite companion. The fact that the primary galaxy is sufficiently apart from its satellite allows us to assume that the tidal phenomena are very small and therefore negligible. In particular, our numerical calculations indicate that the tidal phenomena are significant enough when the distance between the centers of the two galaxies is $R \leq 0.95$. In the present study, we assume that the range of this distance is $1.5 \leq R \leq 3$ and, therefore, the tidal forces can be neglected.

In this rotating frame, the equations of motion are
\begin{eqnarray}
\ddot{x} &=& -\frac{\partial \Phi_{\rm t}}{\partial x} - 2 \Omega_p \dot{y}, \nonumber\\
\ddot{y} &=& -\frac{\partial \Phi_{\rm t}}{\partial y} + 2 \Omega_p \dot{x}, \nonumber\\
\ddot{z} &=& -\frac{\partial \Phi_{\rm t}}{\partial z},
\end{eqnarray}
where the dot indicates derivative with respect to the time. The only integral of motion for the system of differential equations (9) is the well-known Jacobi integral given by the equation
\begin{equation}
J=\frac{1}{2}\left(p_x^2 + p_y^2 + p_z^2 \right) + \Phi_{\rm t}(x,y,z) = E_J,
\end{equation}
where $p_x$, $p_y,$ and $p_z$ are the momenta per unit mass conjugate to $x$, $y,$ and $z$, while $E_J$ is the numerical value of the Jacobi integral.

All numerical outcomes of the present work are based on the numerical integration of the equations of motion (9), which was made using a Bulirsh--St\"{o}er routine in \textsc{fortran}77, with double precision in all subroutines (see Numerical Recipes in FORTRAN, 2nd edn, in Press et al. 1992). The accuracy of the calculations was checked by the consistency of the Jacobi integral (10), which was conserved up to the 12th significant figure.

In this article, we shall use a system of galactic units where the unit of length is 20 kpc, the unit of mass is 1.8 $\times$ $10^{11}$ M$_\odot,$ and the unit of time is 0.99 $\times$ $10^8$ yr. The velocity unit is 197 km\,s\(^{-1}\), while $G$ is equal to unity (see Vozikis \& Caranicolas 1992). In these galactic units, we use the values $a=0.15$, $b=0.368$, $h=0.00625$, $c_{\rm n}=0.0125,$ and $M_{\rm s}=0.2$. The values of the above quantities of the dynamical system remain constant during this research, while the values of $M_{\rm d}$, $M_{\rm n}$, and $R$ are treated as parameters. The above numerical values of the constant dynamical quantities of the system secure positive density everywhere and free of singularities.
\begin{figure*}[!tH]
\centering
\resizebox{\hsize}{!}{\rotatebox{0}{\includegraphics*{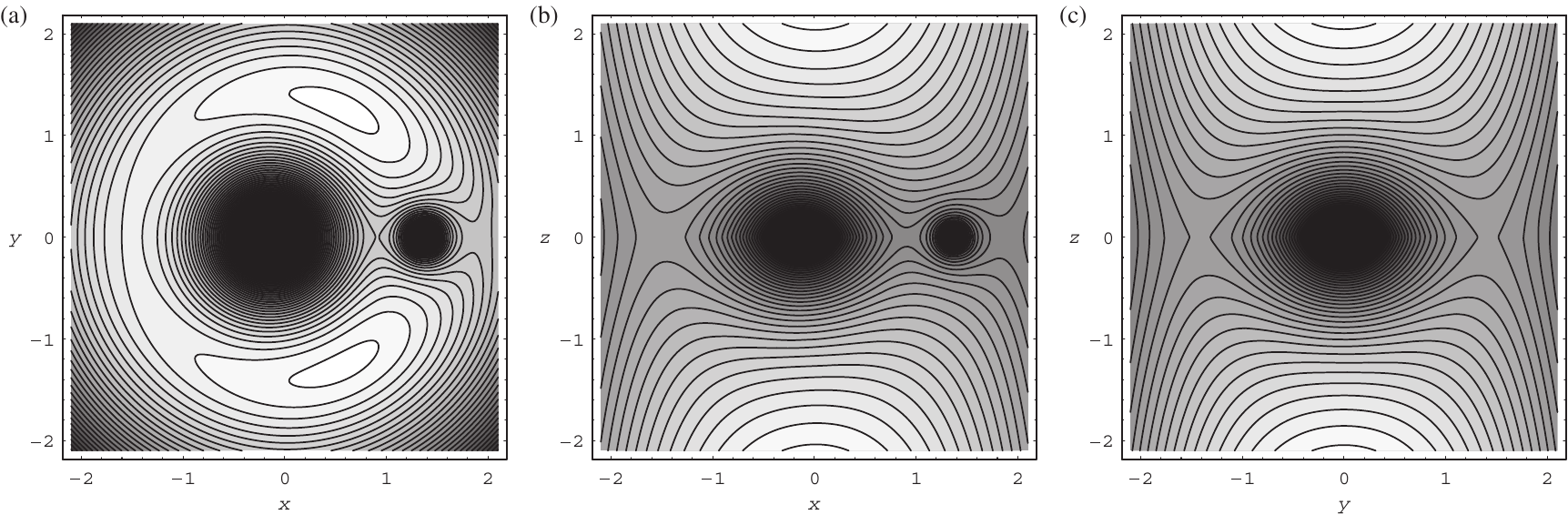}}}
\vskip 0.01cm
\caption{\small (a--c) Contours of the projections of the isopotential curves $\Phi_{\rm t}(x,y,z)=E_J$ on the $(x,y)$, $(x,z),$ and $(y,z)$ planes.}
\end{figure*}
\begin{figure}[!tH]
\centering
\resizebox{\hsize}{!}{\rotatebox{0}{\includegraphics*{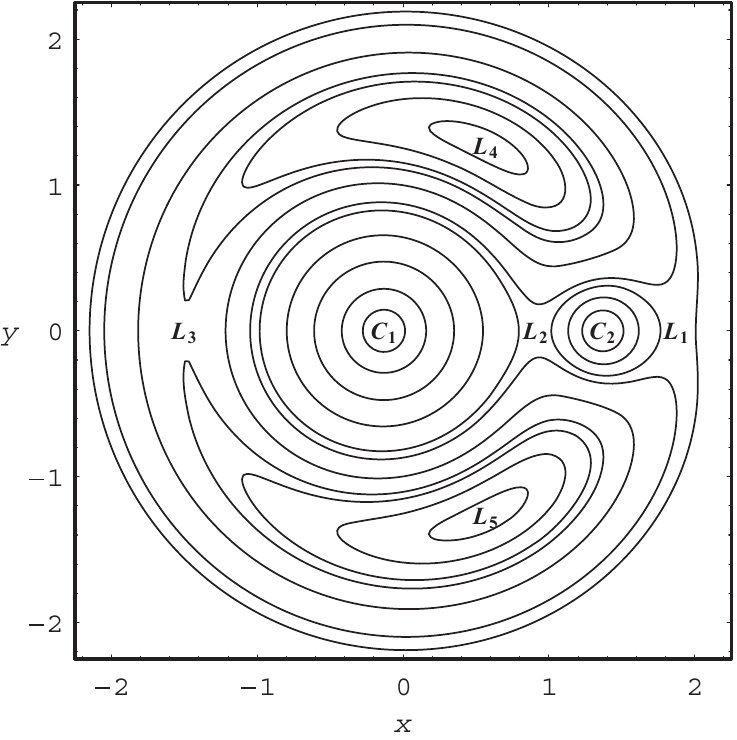}}}
\caption{\small Contours of the projections of the isopotential curves $\Phi_{\rm t}(x,y,z)=E_J$ on the $(x,y)$ plane. The five Lagrange equilibrium points are indicated as $L_1$, $L_2$, $L_3$, $L_4,$ and $L_5$, while $C_1$ and $C_2$ are the centers of the two galaxies at a distance of $R=1.5$.}
\end{figure}
\begin{figure}[!tH]
\centering
\resizebox{\hsize}{!}{\rotatebox{0}{\includegraphics*{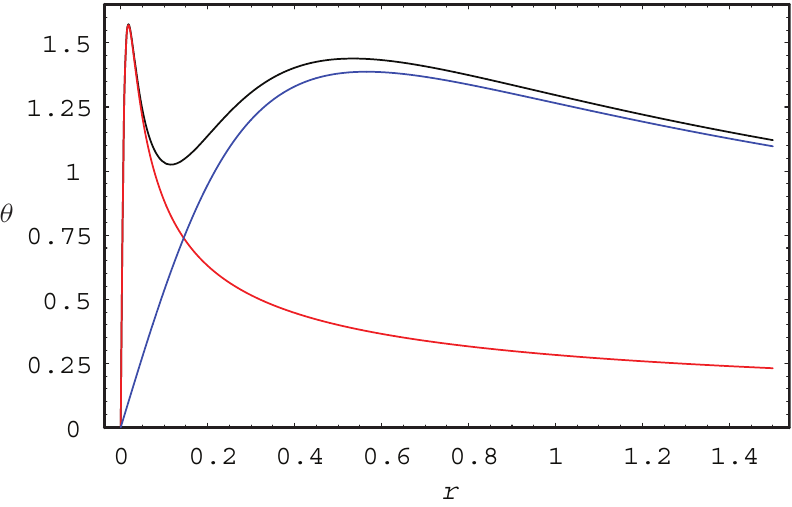}}}
\caption{\small The rotation curve of the primary galaxy is shown as the black line. The red line is the contribution from the spherical nucleus, while the blue line is the contribution from the disk--halo potential. (Colors available only in the electronic version of the article).}
\end{figure}

Figures 1(a--c) show the contours of the projections of the isopotential curves $\Phi_{\rm t}(x,y,z)=E_J$ on the $(x,y)$, $(x,z)$ and $(y,z)$ primary planes, respectively. The values of the parameters are $M_{\rm d}=2$, $M_{\rm n}=0.08$, $R=1.5,$ and $\Omega_p=0.821922$. Lighter colors indicate higher values of $E_J$. Figure 2 shows the contours of the isopotential curves $\Phi_{\rm t}(x,y,z)=E_J$, on the $(x,y)$ plane, with some additional details regarding the stability and the structure of the dynamical system. Moreover, $L_1$, $L_2$, $L_3$, $L_4,$ and $L_5$ are the five Lagrange equilibrium points, while $C_1$ and $C_2$ are the centers of the two galaxies at a distance of $R=1.5$. At these equilibrium points, we have
\begin{equation}
\frac{\partial \Phi_{\rm t}(x,y,z)}{\partial x} = \frac{\partial \Phi_{\rm t}(x,y,z)}{\partial y} = \frac{\partial \Phi_{\rm t}(x,y,z)}{\partial z} = 0.
\end{equation}
Note that $L_1$, $L_2,$ and $L_3$ are the unstable saddle equilibrium points, while $L_4$ and $L_5$ are the triangular points (see Binney \& Tremaine 2008).

The rotation curve of the primary galaxy when $M_{\rm d}=2$ and $M_{\rm n}=0.08$ is shown as the black line in Figure 3. In the same plot, the red line is the contribution from the spherical nucleus, while the blue line is the contribution from the disk--halo component. We observe that at small distances from the primary galactic center $r\leq 0.1$ length units, which equals 2 kpc, dominates the contribution from the spherical nucleus, while at larger distances $r > 2$ kpc, the disk--halo contribution is the dominant factor.
\begin{figure*}[!tH]
\centering
\resizebox{\hsize}{!}{\rotatebox{0}{\includegraphics*{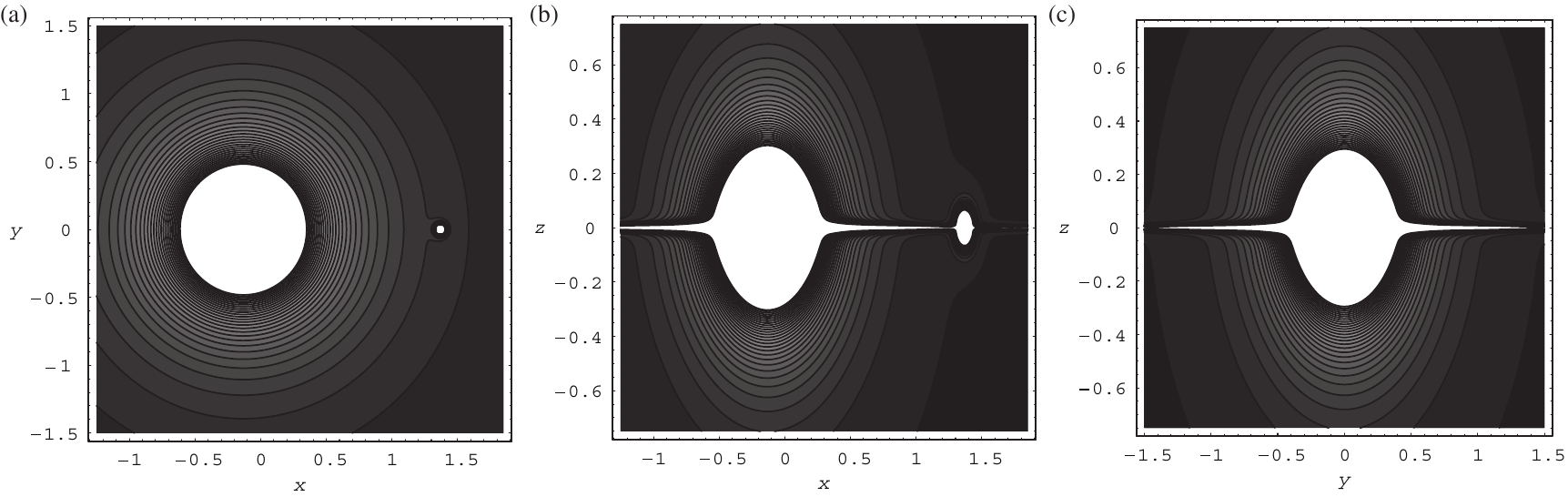}}}
\vskip 0.01cm
\caption{\small (a--c) Contours of the projections of the isodensity curves $\rho(x,y,z)=const$ on the $(x,y)$, $(x,z),$ and $(y,z)$ planes.}
\end{figure*}
\begin{figure*}[!tH]
\centering
\resizebox{\hsize}{!}{\rotatebox{0}{\includegraphics*{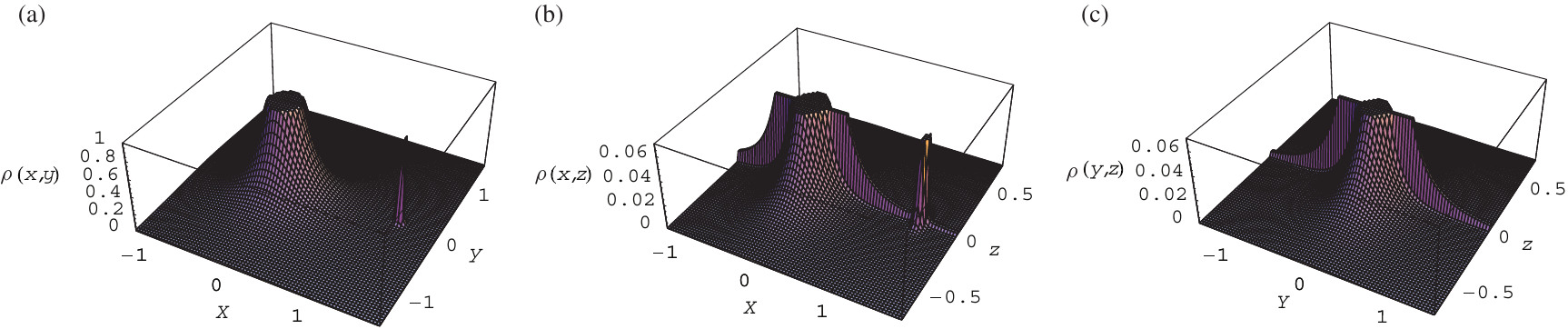}}}
\vskip 0.01cm
\caption{\small (a--c) 3D plots of the distribution of the total mass density $\rho(x,y,z)$ on the $(x,y)$, $(x,z),$ and $(y,z)$ planes.}
\end{figure*}
\begin{figure*}[!tH]
\centering
\resizebox{\hsize}{!}{\rotatebox{0}{\includegraphics*{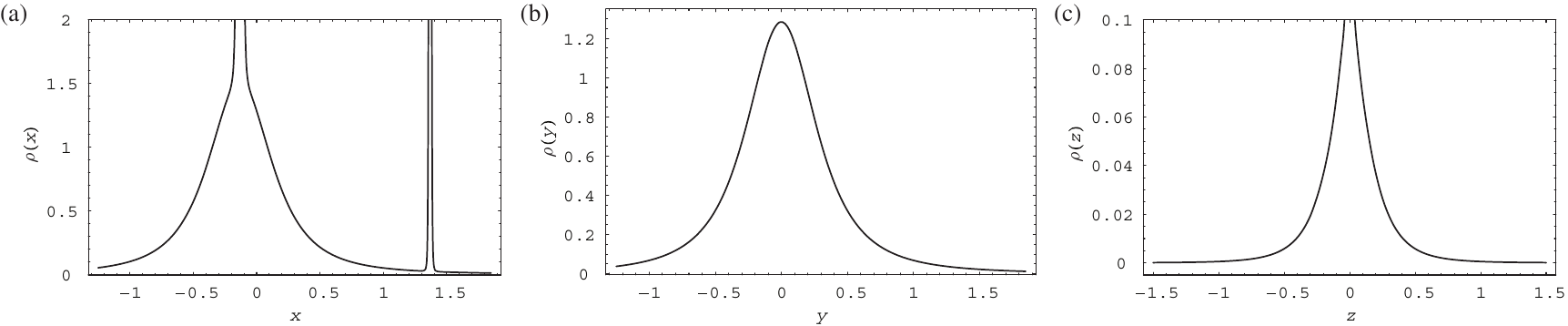}}}
\vskip 0.01cm
\caption{\small (a--c) The evolution of the total mass density along the $x$, $y,$ and $z$ axes.}
\end{figure*}

It would be very illuminating to compute the mass density distribution $\rho(x,y,z)$ of our binary system, derived from the total potential $V_{\rm T}(r,z)$, using the Poisson's equation
\begin{eqnarray}
\rho(x,y,z) &=& \frac{1}{4 \pi} \nabla^2 V_{\rm T}(x,y,z) \nonumber \\
&=& \frac{1}{4 \pi}\left(\frac{\partial^2 V_{\rm T}}{\partial x^2} +\frac{\partial^2 V_{\rm T}}{\partial y^2}
+ \frac{\partial^2 V_{\rm T}}{\partial z^2} \right).
\end{eqnarray}
Figures 4(a--c) show the projections of the isodensity curves $\rho(x,y,z)=const$ on the $(x,y)$, $(x,z),$ and $(y,z)$ primary planes, respectively, when $M_{\rm d}=2$, $M_{\rm n}=0.08$, $R=1.5,$ and $\Omega_p=0.821922$. In the first two plots, we can observe the small variations on density distribution, caused by the mass of the satellite galaxy. Figures 5(a--c) depict the 3D plots of the values of the total mass density $\rho(x,y,z)$ on the same primary planes. Once again, we can distinguish the spherical nucleus at the central region of the primary galaxy, the well-formed disk structure, and also the abrupt peak of the mass density at the position of the satellite galaxy $(x \simeq 1.36)$. These 3D plots ensure that the mass density of our binary galactic system has positive values everywhere. Going one step further, we present in Figures 6(a--c) the evolution of the mass density along the $x$, $y,$ and $z$ axes, respectively. One can observe that in all cases the mass density obtains high values near the two galactic centers, while it reduces and tends asymptotically to zero, as the distance from the centers of the two galaxies increases. The values of the mass density $\rho$ on the vertical axis are in M$_{\odot}$\,pc$^{-3}$.

\section{Structure of the 2D Hamiltonian System}

In this section, we shall investigate the properties of motion in the Hamiltonian system of two degrees of freedom. This can be derived from Equation (10) if we set $z=p_z=0$. Then, the corresponding Hamiltonian is
\begin{equation}
J_2 = \frac{1}{2}\left(p_x^2 + p_y^2 \right) + \Phi_{\rm t}(x,y) = E_{J2},
\end{equation}
where $E_{J2}$ is the numerical value of $J_2$. As the dynamical system is now two-dimensional, we can use the classical, qualitative method of plotting the successive intersections of the 2D orbits, using the $(x, p_x)$, $y=0$, $p_y > 0$ Poincar\'{e} surface of section (PSS), in order to determine the regular or chaotic character of motion. This method has been extensively applied to Hamiltonian systems with two degrees of freedom, as in these systems the PSS is a two-dimensional plane. The results obtained from the study of the 2D system will be used to help us understand the structure of the more complicated phase space of the 3D system, which will be presented in the following section.
\begin{figure*}[!tH]
\centering
\resizebox{\hsize}{!}{\rotatebox{0}{\includegraphics*{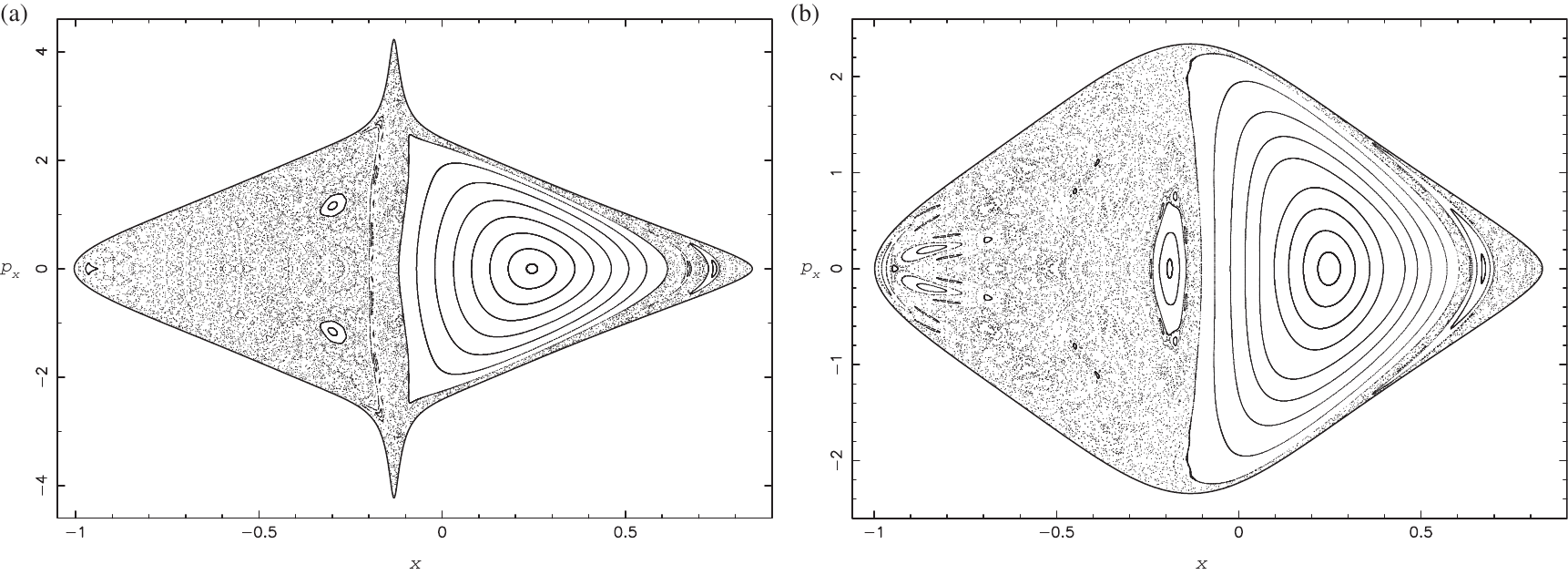}}}
\vskip 0.01cm
\caption{\small The $(x, p_x)$ Poincar\'{e} phase plane when $R=1.5$ and $\Omega_p=0.821922$. (a) $M_{\rm d}=2$ and $M_{\rm n}=0.08$; (b) $M_{\rm d}=2.08$ and $M_{\rm n}=0$.}
\end{figure*}
\begin{figure*}[!tH]
\centering
\resizebox{\hsize}{!}{\rotatebox{0}{\includegraphics*{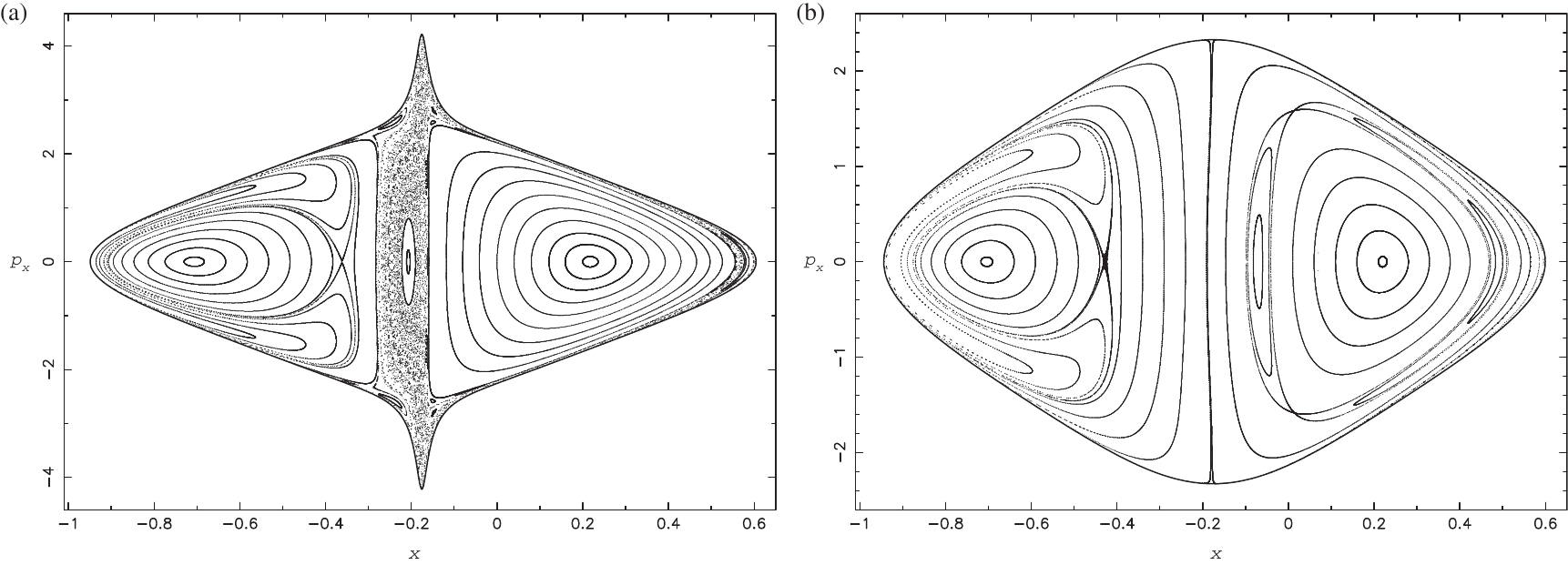}}}
\vskip 0.01cm
\caption{\small (a--b) Similar to Figures 7(a--b) but when $R=2$ and $\Omega_p=0.533854$.}
\end{figure*}
\begin{figure*}[!tH]
\centering
\resizebox{\hsize}{!}{\rotatebox{0}{\includegraphics*{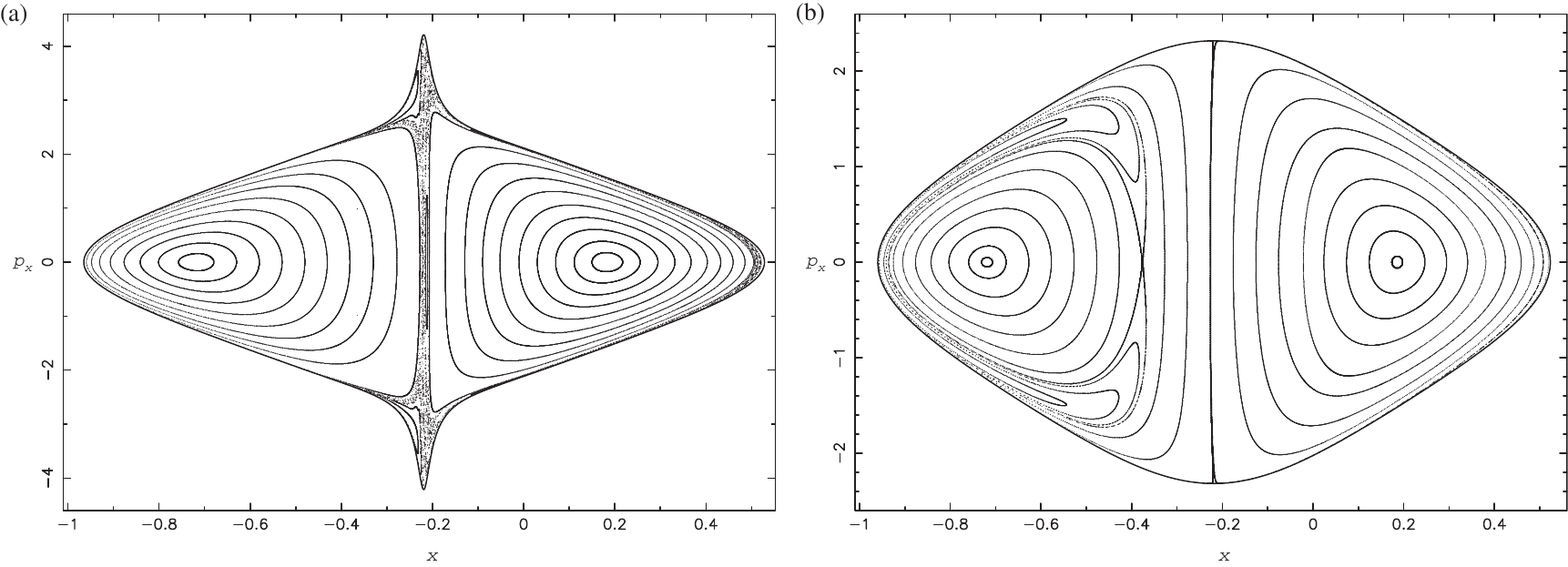}}}
\vskip 0.01cm
\caption{\small (a--b) Similar to Figures 7(a--b) but when $R=2.5$ and $\Omega_p=0.381995$.}
\end{figure*}
\begin{figure*}[!tH]
\centering
\resizebox{\hsize}{!}{\rotatebox{0}{\includegraphics*{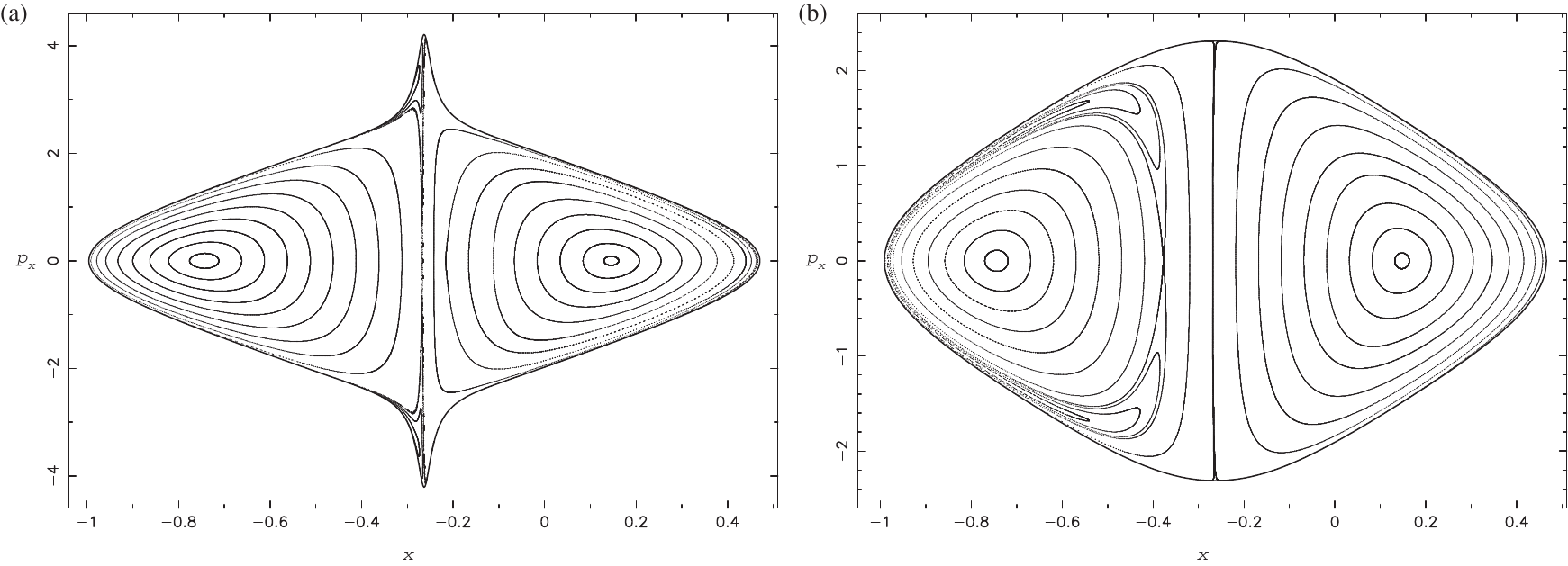}}}
\vskip 0.01cm
\caption{\small (a--b) Similar to Figures 7(a--b) but when $R=3$ and $\Omega_p=0.290593$.}
\end{figure*}

Figures 7(a--b) show the $(x, p_x)$, $y=0$, $p_y>0$ Poincar\'{e} phase plane, for the motion of a star in the primary galaxy, which obtained by numerical integration of the equations of motion (9). In this case, the distance between the centers of the two galaxies is $R=1.5$ and $\Omega_p=0.821922$. The value of the Jacobi integral is $E_{J2}=-2.6$. This particular value of the Jacobi integral will remain constant for all our numerical experiments. In Figure 7(a) we have $M_{\rm d}=2$ and $M_{\rm n}=0.08$, that is, the primary galaxy has a dense and massive nucleus. We observe a large chaotic area, while the regular region is mainly confined around the stable retrograde periodic point. There are also some small islands of invariant curves embedded in the chaotic sea, corresponding to secondary resonant orbits. Note that near the center of the primary galaxy, the velocity obtains high values of the order of 900 km\,s\(^{-1}\). This results from the presence of the dense nucleus in the core of the galaxy and it is characteristic of strong galactic activity. The outermost black solid line defines the zero velocity curve. Figure 7(b) is similar to Figure 7(a), but when $M_{\rm d}=2.08$ and $M_{\rm n}=0$. As the total mass of the system is conserved, this means that now the disk occupies all the available mass of the primary galaxy. In this case, the pattern has two main differences from the pattern shown in Figure 7(a). The first difference is that the chaotic region is smaller and the second is that the velocity near the galactic core is smaller, about 470 km\,s\(^{-1}\), than that of Figure 7(a). Moreover, some secondary resonances appear in the area of the retrograde orbits. Some small sticky regions are also present in both cases (active or quiet galaxy). Therefore, one can say that the structure of chaos in galaxies is a result not only of galactic interaction but also of the nuclear galactic activity. In other words, the presence of the dense and massive nucleus in the galactic core of the primary galaxy affects drastically the character of motion of the stars.

Figure 8(a) is similar to Figure 7(a) but when the distance between the centers of two the galaxies is $R=2$ and $\Omega_p=0.533854$. The majority of the orbits around the direct and retrograde periodic point are ordered. There is a significant chaotic area mainly near the central region of the galaxy. A high-velocity value is once again observed near the galactic core of the primary galaxy, while some secondary resonances are also present. Figure 8(b) is similar to Figure 8(a), but when the galaxy is quiet, that is, when $M_{\rm n}=0$. Here, the secondary resonances look more prominent, while only a small and confined chaotic layer around the direct periodic point is present. Furthermore, as the primary galaxy is quiet, the velocity near the center of the galactic core is smaller than that observed in Figure 8(a). Figures 9(a--b) are similar to Figures 8(a--b), but when the distance between the centers of the two galaxies is $R=2.5$ and $\Omega_p=0.381995$. In Figure 9(a), we observe that there is a relatively small chaotic layer confined mainly near the central region of the galaxy, while the rest of the phase plane is covered by invariant curves corresponding to regular orbits circulating around the stable direct and retrograde periodic point. Figure 9(b) is similar to Figure 9(a) when the galaxy is quiet. All orbits seem to be regular. Chaotic motion was not observed and, if present, is negligible. Figures 10(a--b) are similar to Figures 9(a--b) but when the distance between the centers of the two galaxies is $R=3$ and $\Omega_p=0.290593$. In Figure 10(a), we observe that the entire phase plane is covered by invariant curves corresponding to regular orbits circulating around the stable direct and retrograde periodic point, while there is no indication of chaotic motion. Figure 10(b) is similar to Figure 10(a), but when the galaxy is quiet. Once again, all orbits are regular and chaotic motion was not observed. Therefore, we may conclude that the primary galaxy, for large separations from its companion, does not show chaotic motion irrespective of the presence of the  dense and massive nucleus in the galactic core. The outermost solid curve in the $(x, p_x)$ phase planes shown in Figures 7--10 is the limiting curve in each case.
\begin{figure}[!tH]
\centering
\resizebox{\hsize}{!}{\rotatebox{0}{\includegraphics*{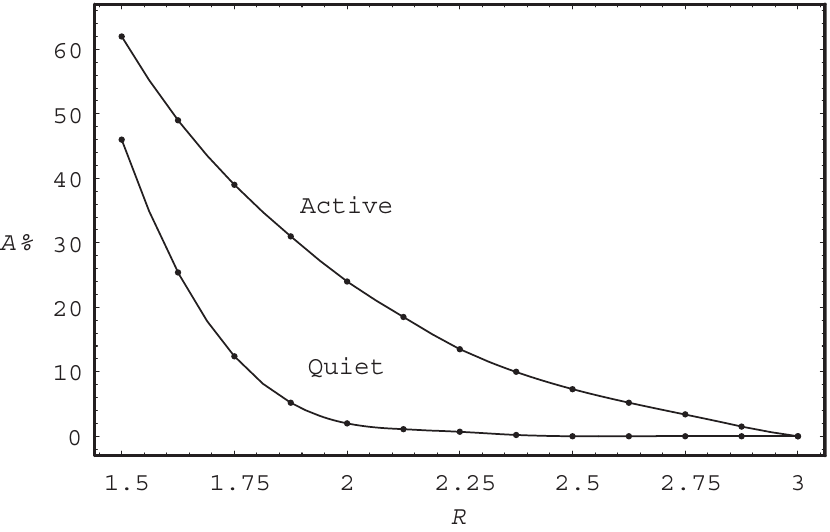}}}
\caption{\small A plot of the area $A\%$ of the $(x,p_x)$ phase plane covered by chaotic orbits as a function of the distance $R$, when the primary galaxy is active or quiet.}
\end{figure}

Figure 11 shows the percentage $A\%$ on the $(x, p_x)$ phase plane covered by chaotic orbits as a function of the distance $R$ between the centers of the two galaxies, in two different cases. In the first case, the primary galaxy is active and the values of the parameters are $M_{\rm d}=2$ and $M_{\rm n}=0.08$, while in the second case the primary galaxy is quiet and, therefore, the values of the parameters are $M_{\rm d}=2.08$ and $M_{\rm n}=0$. The value of the angular frequency $\Omega_p$ is calculated for each particular value of the distance $R$ from Equation (3). The range of the values regarding the distance between the centers of the two galaxies is $1.5 \leq R \leq 3$. We observe in Figure 11 that the chaotic percentage $A\%$ decreases exponentially, as the distance $R$ increases in both cases (active and quiet primary galaxy). Moreover, both fitting curves converge to zero, when $R=3$. The lower fitting curve, which corresponds to the case when the primary galaxy is quiet, presents a more rapid reduction and reaches more quickly (when $R \geq$ 2.35) to the zero limit. A more detailed view of Figure 11 reveals that for each particular value of the distance $R$, the chaotic percentage $A\%$ is always smaller when the primary galaxy is quiet. We must point out that the chaotic percentage $A\%$ is calculated as follows: we choose $10^3$ orbits with random initial conditions $(x_0, p_{x0})$ in each phase plane and then divide the number of those who correspond to chaotic orbits to the total number of the tested orbits.
\begin{figure}[!tH]
\centering
\resizebox{\hsize}{!}{\rotatebox{0}{\includegraphics*{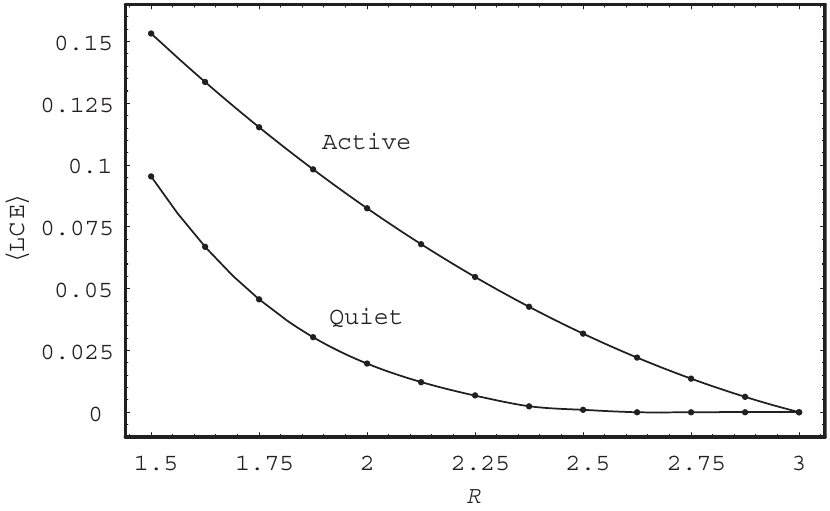}}}
\caption{\small A plot of the average value of the  $\langle {\rm LCE}\rangle$ as a function of the distance $R$, when the primary galaxy is active or quiet, for the 2D dynamical system.}
\end{figure}

\begin{figure*}
\centering
\resizebox{0.80\hsize}{!}{\rotatebox{0}{\includegraphics*{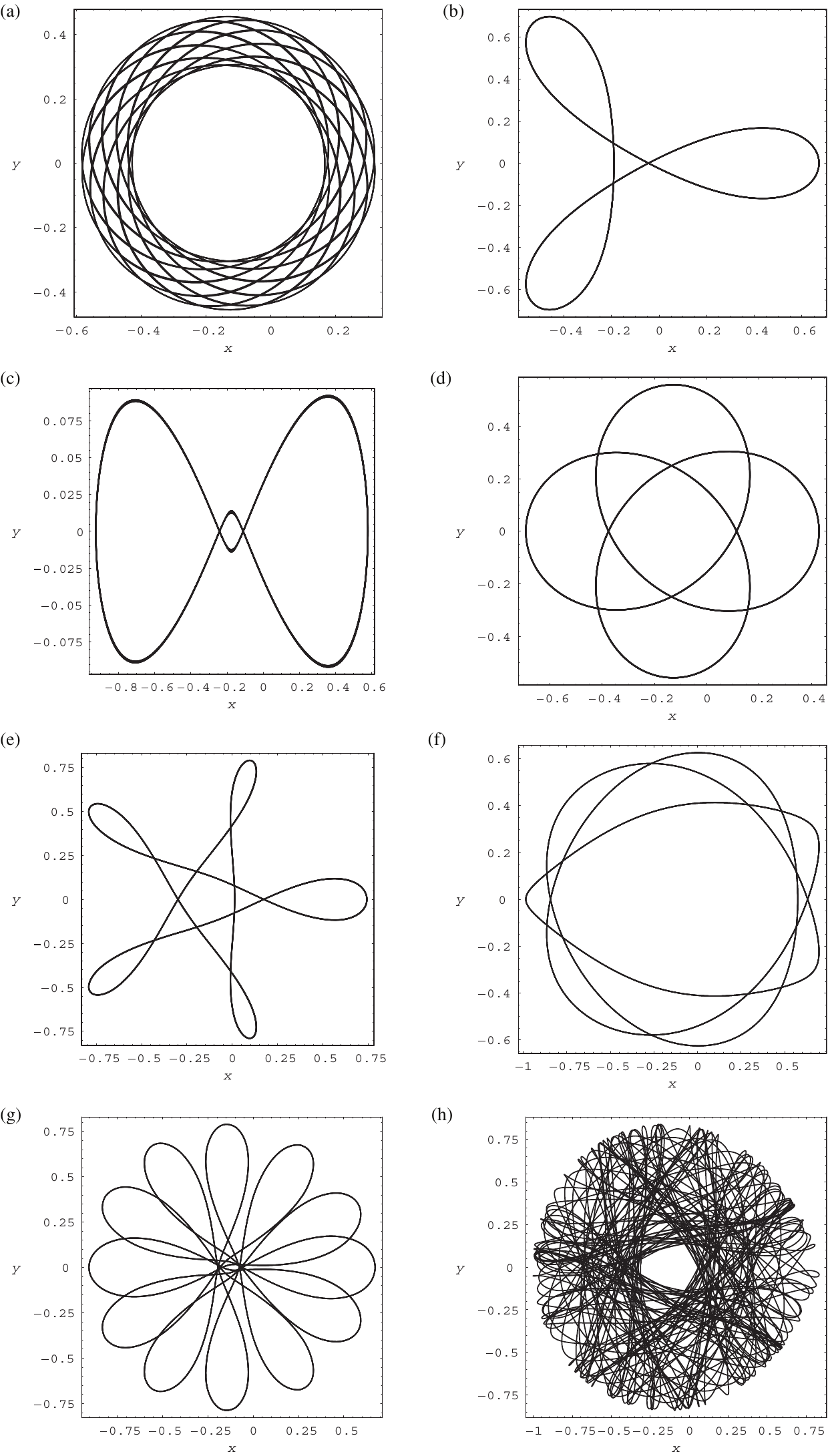}}}
\vskip 0.01cm
\caption{\small (a--h) Eight representative regular orbits of the 2D dynamical system. The values of the initial conditions and all the other parameters are given in the text.}
\end{figure*}

To have a better estimation about the degree of chaos in our dynamical system in every case, we have computed the average value of the Lyapunov characteristic exponent (LCE; see Lichtenberg \& Lieberman 1992). Figure 12 shows a plot of the $\langle {\rm LCE}\rangle$ as a function of the distance $R$ between the centers of the two galaxies, in two different cases. As previously, in the first case, the primary galaxy is active and the values of the parameters are $M_{\rm d}=2$ and $M_{\rm n}=0.08$, while in the second case the primary galaxy is quiet and, therefore, the values of the parameters are $M_{\rm d}=2.08$ and $M_{\rm n}=0$. The value of the angular frequency $\Omega_p$ is calculated for each particular value of the distance $R$ from Equation (3). The range of values regarding the distance between the centers of the two galaxies is again $1.5 \leq R \leq 3$. One can see in Figure 12 that $\langle {\rm LCE}\rangle$ decreases almost linearly as the distance $R$ increases when the primary galaxy is active. On the contrary, when the nucleus is absent, we observe that the corresponding fitting curve presents a more rapid reduction. Once again, as pointed out in Figure 11, for each particular value of the distance $R$, the average value $\langle {\rm LCE}\rangle$ is always smaller when the primary galaxy is quiet. Moreover, both fitting curves converge to zero when $R=3$. Here we must note that it is well known that the value of the LCE is different in each chaotic component (see Saito \& Ichimura 1979). As we have in all cases regular regions and only one unified chaotic area in each $(x,p_x)$ phase plane, we calculated the average value of the LCE by taking $10^3$ orbits with random initial conditions $(x_0, p_{x0})$ in the chaotic area in each case and we integrated each orbit for a time interval of $10^5$ time units in order to obtain a reliable value of the LCE regarding the nature of the orbit. Here we have to point out that all calculated LCEs corresponding to chaotic orbits were different only on the fifth decimal point in the same chaotic region.

In Figures 13(a--h) we present eight representative orbits of the 2D dynamical system. Figure 13(a) shows a quasi-periodic orbit circulating around the primary galaxy, with initial conditions $x_0=0.167$, $y_0=0$, and $p_{x0}=0$, while the value of $p_{y0}$ is obtained from the Jacobi integral (13) for all orbits. The values of all the other parameters are as in Figure 7(a). Figure 13(b) shows a periodic orbit moving around the center of the primary galaxy, with initial conditions $x_0=-0.1895$, $y_0=0$, and $p_{x0}=0$, while the values of all other parameters are as in Figure 7(b). This orbit is characteristic of the 1:2 resonance. In Figure 13(c), a figure-eight type periodic orbit, characteristic of the 1:3 resonance, is shown. The initial conditions are $x_0=0.5788$, $y_0=0$, and $p_{x0}=0$, while the value of all other parameters are as in Figure 8(a). Figure 13(d) shows a periodic orbit moving around the primary galaxy with initial conditions $x_0=0.4295$, $y_0=0$, and $p_{x0}=0$, while the values of all the other parameters are as in Figure 7(b). This orbit is a characteristic example of the 3:3 resonance. In Figure 13(e), a periodic orbit with initial conditions $x_0=0.7445$, $y_0=0$, and $p_{x0}=0$ is presented. This orbit belongs to the family of the 2:3 resonant orbits. The values of all other parameters are as in Figure 7(a). In Figure 13(f), we see a periodic orbit with initial conditions $x_0=-0.9854$, $y_0=0$, and $p_{x0}=0$. The values of all other parameters for this orbit are as in Figure 7(b). Figure 13(g) shows a complicated resonant periodic orbit of higher multiplicity, circulating around the primary galaxy, with initial conditions $x_0=0.673$, $y_0=0$, and $p_{x0}=0$, while the values of all the other parameters are as in Figure 7(a). This orbit produces a set of nine tiny islands of invariant curves which are embedded in the retrograde area of the unified chaotic domain shown in Figure 7(a). A chaotic orbit with initial conditions $x_0=-0.31$, $y_0=0$, and $p_{x0}=0$ is given in Figure 13(h). The values of all other parameters for this orbit are as in Figure 7(a). We observe that stars moving in chaotic orbits pass arbitrarily around the galactic core without getting close enough. All orbits shown in Figures 13(a--h) were calculated for a time period of 150 time units.

\section{Structure of the 3D Hamiltonian System}

In this section, we shall investigate the regular or chaotic nature of motion in the 3D Hamiltonian system described by Equation (10). In order to keep things simple, we shall use our experience gained from the study of the 2D dynamical system in order to obtain a clear picture regarding the properties of motion in the 3D dynamical model. We are particularly interested in locating the initial conditions in the 3D dynamical system, producing regular or chaotic orbits. A convenient way to obtain this is to start from the $(x, p_x)$ phase planes of the 2D system with the same value of the Jacobi integral as used in the 2D system and described in the previous section. Specifically, the regular or chaotic nature of the 3D orbits is found as follows: we choose initial conditions $\left(x_0, p_{x0}, z_0\right)$, $y_0=p_{z0}=0$, such that $\left(x_0, p_{x0}\right)$ is a point on the phase planes of the 2D system. The points $\left(x_0, p_{x0}\right)$ lie inside the limiting curve
\begin{equation}
\frac{1}{2}p_x^2 + \Phi_{\rm t}(x) = E_{J2},
\end{equation}
which is the limiting curve containing all the invariant curves of the 2D system. Thus, we take $E_J=E_{J2}$. For this purpose, a large number of orbits (about $10^3$) were computed with initial conditions $\left(x_0, p_{x0},z_0\right)$, where $\left(x_0, p_{x0}\right)$ is a point in the chaotic regions of the $(x,p_x)$ phase planes of Figures 7(a--b), 8(a--b), 9(a--b), and 10(a--b), with all permissible values of $z_0$ and $p_{z0}=0$. Remember that as we are on the phase plane, we have $y_0=0$, while in all cases the value of $p_{y0}$ was obtained from the Jacobi integral (10). All tested orbits were found to be chaotic. Moreover, it is well known that, usually, chaotic domains in 2D systems lead to chaotic motion in 3D space when an additional degree of freedom is added, like a `perturbation'. Therefore, one may conclude that orbits with the initial conditions $\left(x_0, p_{x0}\right)$ be a point in the chaotic regions of the 2D phase planes and, for all permissible values of $z_0$, remain chaotic in the 3D system as well.
\begin{figure}[!tH]
\centering
\resizebox{\hsize}{!}{\rotatebox{0}{\includegraphics*{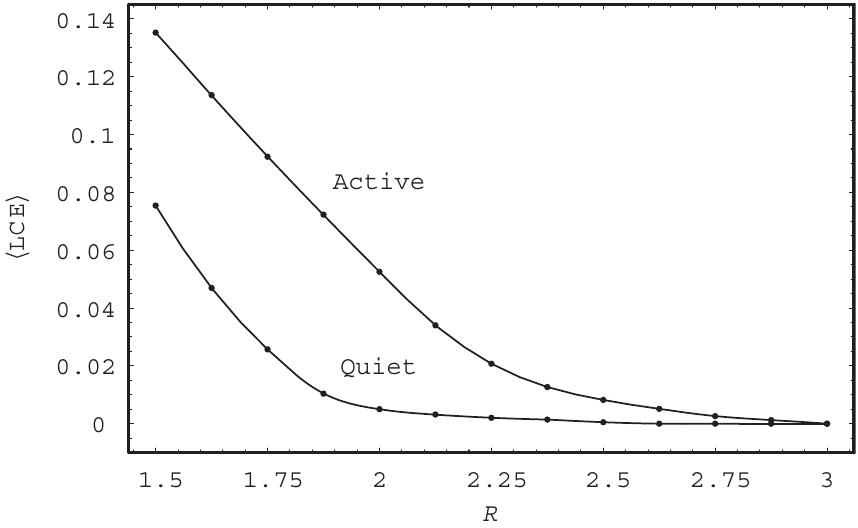}}}
\caption{\small A plot of the average value of the maximum $\langle {\rm LCE}\rangle$ as a function of the distance $R$, when the primary galaxy is active or quiet, for the 3D dynamical system.}
\end{figure}

In Figure 14, we present a plot of the average value of the LCE of the 3D system, as a function of the distance $R$ between the centers of the two galaxies, in two different cases. As in the previous section, in the first case, the primary galaxy is active and the values of the parameters are $M_{\rm d}=2$ and $M_{\rm n}=0.08$, while in the second case the primary galaxy is quiet and, therefore, the values of the parameters are $M_{\rm d}=2.08$ and $M_{\rm n}=0$. The value of the angular frequency $\Omega_p$ is calculated for each particular value of the distance $R$ from Equation (3). The range of values regarding the distance between the centers of the two galaxies is again $1.5 \leq R \leq 3$. One can see in Figure 14 that the $\langle {\rm LCE}\rangle$ decreases almost linearly as the distance $R$ increases when the primary galaxy is active when $R \leq 2.25$. On the contrary, when the nucleus is absent, we observe that the corresponding fitting curve presents a more rapid reduction. Once again, as pointed out in Figure 12 regarding the 2D system, for each particular value of the distance $R$, the average value $\langle {\rm LCE}\rangle$ is always smaller when the primary galaxy is quiet. Moreover, both fitting curves converge earlier to zero value, than in Figure 12. The method we used in order to compute the $\langle {\rm LCE}\rangle$ in each case is the same as described in Figure 12. For all tested chaotic orbits in the 3D system, the initial value of $z_0$ is common and equal to 0.1. Here we must point out that if we compare the plot of the 3D system shown in Figure 14 with that of the 2D system shown in Figure 12, we observe that in each case (active and quiet galaxy, respectively) the average values of the LCEs of the 3D system are smaller than those of the 2D system. Therefore, we may conclude that the degree of chaos in the primary galaxy is smaller when the dynamical system has three degrees of freedom.

Our next step is to study the character of orbits with initial conditions $\left(x_0, p_{x0}, z_0\right)$, $y_0=p_{z0}=0$, such that $\left(x_0, p_{x0}\right)$ is a point in the regular regions of Figures 7(a--b), 8(a--b), 9(a--b), and 10(a--b). The phase space of a conservative system of three degrees of freedom has six dimensions, i.e. in Cartesian coordinates $\left(x, y, z, \dot{x}, \dot{y}, \dot{z}\right)$. For a given value of the Jacobi integral, a trajectory lies on a five-dimensional manifold. In this manifold, the surface of the section is four-dimensional. This does not allow us to visualize and interpret directly the structure and the properties of the phase space in dynamical systems of three degrees of freedom. One way to overcome this problem is to project the surface of the section to space with lower dimensions. In fact, here, we will apply the method introduced by Pfenniger (1984) (see also Revaz \& Pfenniger 2001). We take sections in the plane $y=0$, $p_y >0$ of 3D orbits, whose initial conditions differ from the plane parent periodic orbits only by the $z$ component. The set of the resulting four-dimensional points in $\left(x, p_x, z, p_z\right)$ phase space is projected on the $\left(z, p_z\right)$ plane. If the projected points lie on a well-defined curve, we call it an `invariant curve', then the motion is regular, while if not, the motion is chaotic. The projected points on the $\left(z, p_z\right)$ plane show nearly invariant curves around the periodic points at $z=0$, $p_z=0$, as long as the coupling is weak. When the coupling is stronger, the corresponding projections in the $\left(z, p_z\right)$ plane displays an increasing departure of the plane periodic point, up to making a direct orbit a retrograde one and vice versa. Here, we must define what one means by direct and retrograde 3D orbit. If consequents in the $\left(z, p_z\right)$ section of the 3D orbit drop in one of the two domains of the corresponding section of 2D orbits at the same value of the Jacobi integral $E_J$, we can distinguish between direct and retrograde motion. Orbits which visit both domains are intermittently direct or retrograde.
\begin{figure*}[!tH]
\centering
\resizebox{\hsize}{!}{\rotatebox{0}{\includegraphics*{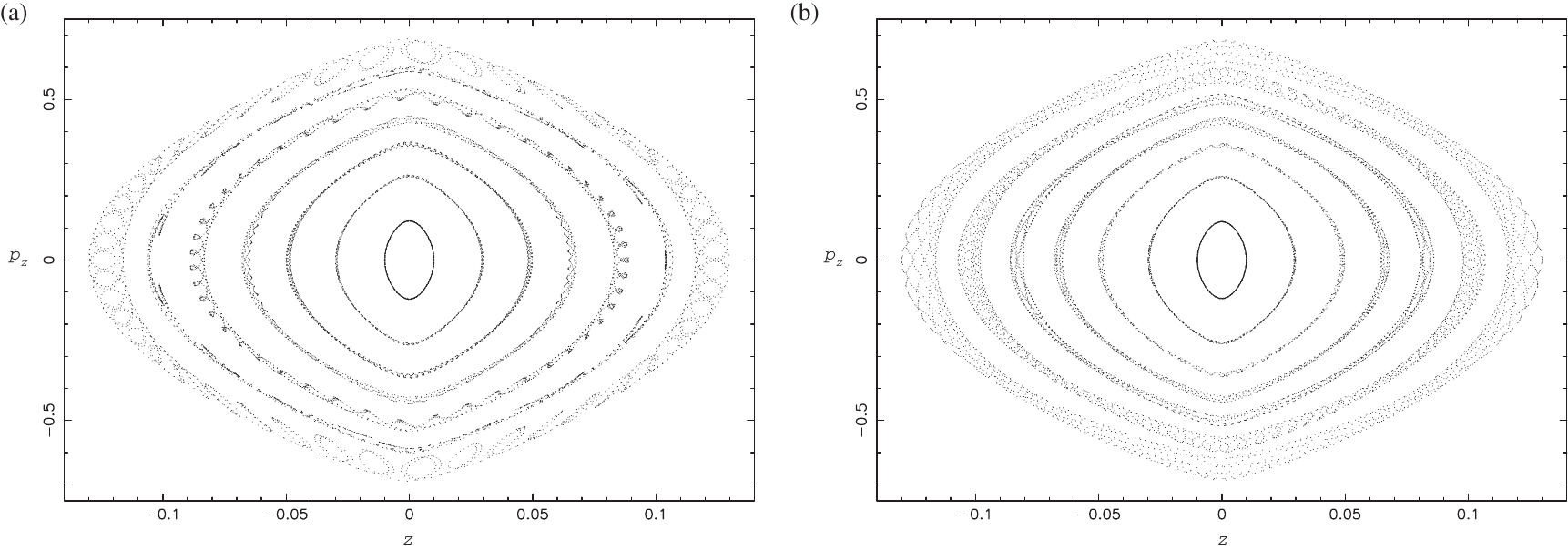}}}
\vskip 0.01cm
\caption{\small (a--b) Projections of the sections of 3D orbits with the plane $y=0$ when $p_y>0$. The set of the four-dimensional points $\left(x, p_x, z, p_z\right)$ is projected on the $\left(z, p_z\right)$ plane.}
\end{figure*}

Figures 15(a--b) show two typical $\left(z, p_z\right)$ sections of 3D orbits, starting with initial conditions close to two different stable periodic points on the $\left(x, p_x\right)$ phase planes of the 2D system. In order to obtain the results shown in Figure 15(a), we have taken the point $\left(x_0, p_{x0}\right)=(0.25, 0)$ representing approximately the position of the retrograde periodic point in the $\left(x, p_x\right)$ phase plane of Figure 7(a). Similarly, in Figure 15(b) we can observe the $\left(z, p_z\right)$ projections near the retrograde periodic point in the $\left(x, p_x\right)$ phase plane of Figure 7(b). The position of the periodic point is $\left(x_0, p_{x0}\right)=(0.25, 0)$. Note that in both cases the numerical results indicate that for small values of $z_0$ the motion is regular, while for larger values of $z_0$ the motion becomes chaotic. The integration time for each `invariant curve' shown in Figures 15(a--b) is 3 $\times$ $10^3$ time units. Numerical calculations not given here suggest that the above method can be applied in all regular regions around each retrograde or direct stable periodic point of Figures 7(a--b), 8(a--b), 9(a--b), and 10(a--b).

We must emphasize that the results presented in Figures 15(a--b) are rather qualitative and can be considered as an indication that the transition from regularity to chaos in 3D orbits occurs as the value of $z_0$ increases. In order to form a more complete and accurate view of the phase space in the 3D system, we computed a large number of 3D orbits (approximately $10^3$) near each periodic point of the $\left(x, p_x\right)$ phase planes of the 2D system, for different initial conditions $\left(x_0, p_{x0}\right)$ and also for different values of $z_0$. Our target was to determine the average minimum value of $z_0$ for which the nature of a 3D orbit changes from regular to chaotic. Table 1 shows the value $\langle z_{\rm min}\rangle$ near the direct and retrograde stable periodic points of Figures 7(a--b), 8(a--b), 9(a--b), and 10(a--b), for four different values of distance $R$ between the centers of the two galaxies. From Table 1 we can induce three important results. (i) In every case (active or quiet primary galaxy) the minimum value of $z_0$ in the regions near the retrograde stable periodic points is always larger than the corresponding of the regions of the direct periodic points. (ii) If we compare each kind of galaxy regarding the nuclear activity, (active or quiet) for the same distance $R$, we observe that when the nucleus is not present at the galactic core (quiet -- non-active primary galaxy), the 3D orbits can approach higher values of $z_0$ and remain regular. On the other hand, when the nucleus is present, the value of $\langle  z_{\rm min}\rangle$ is smaller. (iii) As the distance between the centers of the galaxies increases, the minimum initial value of $z_0$ for which a 3D can remain regular increases in both cases (active or quiet primary galaxy). This means that as the two galaxies are at large distances, their mutual interactions are weak enough and, therefore, the majority of the 3D orbits are ordered. Moreover, at large distances, one can conclude that the main factor responsible for the observed chaotic motion is the nuclear activity of the nucleus in the core of the primary galaxy.
\begin{table}[ht]
\begin{center}
\caption{Average Value of the Minimum $z_0$ Near the Direct and Retrograde Periodic Points, for Four Different Values of the Distance $R$}
\ \ \ \\
\begin{tabular}{|c|c|c||c|}
\hline
{Distance} & {Case} & {Region} & $\langle  z_{\rm min}\rangle$ \bigstrut[t] \\
\hline \hline
\multirow{4}[4]{*}{$R=1.5$} &
\multirow{2}[4]{*}{Active}
          & Direct
                       &   --    \bigstrut[t] \\ \cline{3-4}
&         & Retrograde & 0.0792 \bigstrut[t] \\ \cline{2-4} &
\multirow{2}[4]{*}{Quiet}
          & Direct
                       &   --    \bigstrut[t] \\ \cline{3-4}
&         & Retrograde & 0.0802 \bigstrut[t] \\ \cline{2-4}
\hline
\multirow{4}[4]{*}{$R=2.0$} &
\multirow{2}[4]{*}{Active}
          & Direct
                       & 0.0865 \bigstrut[t] \\ \cline{3-4}
&         & Retrograde & 0.0912 \bigstrut[t] \\ \cline{2-4} &
\multirow{2}[4]{*}{Quiet}
          & Direct
                       & 0.0896 \bigstrut[t] \\ \cline{3-4}
&         & Retrograde & 0.0964 \bigstrut[t] \\ \cline{2-4}
\hline
\multirow{4}[4]{*}{$R=2.5$} &
\multirow{2}[4]{*}{Active}
          & Direct
                       & 0.0983 \bigstrut[t] \\ \cline{3-4}
&         & Retrograde & 0.1025 \bigstrut[t] \\ \cline{2-4} &
\multirow{2}[4]{*}{Quiet}
          & Direct
                       & 0.1012 \bigstrut[t] \\ \cline{3-4}
&         & Retrograde & 0.1107 \bigstrut[t] \\ \cline{2-4}
\hline
\multirow{4}[4]{*}{$R=3.0$} &
\multirow{2}[4]{*}{Active}
          & Direct
                       & 0.1098 \bigstrut[t] \\ \cline{3-4}
&         & Retrograde & 0.1157 \bigstrut[t] \\ \cline{2-4} &
\multirow{2}[4]{*}{Quiet}
          & Direct
                       & 0.1185 \bigstrut[t] \\ \cline{3-4}
&         & Retrograde & 0.1214 \bigstrut[t] \\ \cline{2-4}
\hline
\end{tabular}
\end{center}
\end{table}

So far, we have seen that 3D orbits with initial conditions $\left(x_0, p_{x0}, z_0\right)$, such that $\left(x_0, p_{x0}\right)$ is a point in the chaotic regions of the 2D system, for all permissible values of $z_0$ are chaotic. On the other hand, the nature (ordered or chaotic) of 3D orbits with initial conditions $\left(x_0, p_{x0},z_0\right)$, such that $\left(x_0, p_{x0}\right)$ is a point in the regular regions around the stable direct or retrograde periodic points of the 2D system, depends on the particular value of $z_0$, as we can see in Table 1. We did not feel that it was necessary to try to define the values of $\langle  z_{\rm min}\rangle$ for each regular region of the 2D system corresponding to secondary resonances which are represented by sets of multiple small islands of invariant curves in the $(x, p_x)$ phase planes shown in Figures 7(a--b), 8(a--b), 9(a--b), and 10(a--b). Numerical results indicate that 3D orbits with initial conditions $\left(x_0, p_{x0}, z_0\right)$, such that $\left(x_0, p_{x0}\right)$ is a point in the regular regions corresponding to secondary resonances of the 2D system, remain regular for $\langle  z_{\rm min}\rangle \lesssim 0.052$, while for larger values of $z_0$, they change their character from regular to chaotic.
\begin{figure*}[!tH]
\centering
\resizebox{0.75\hsize}{!}{\rotatebox{0}{\includegraphics*{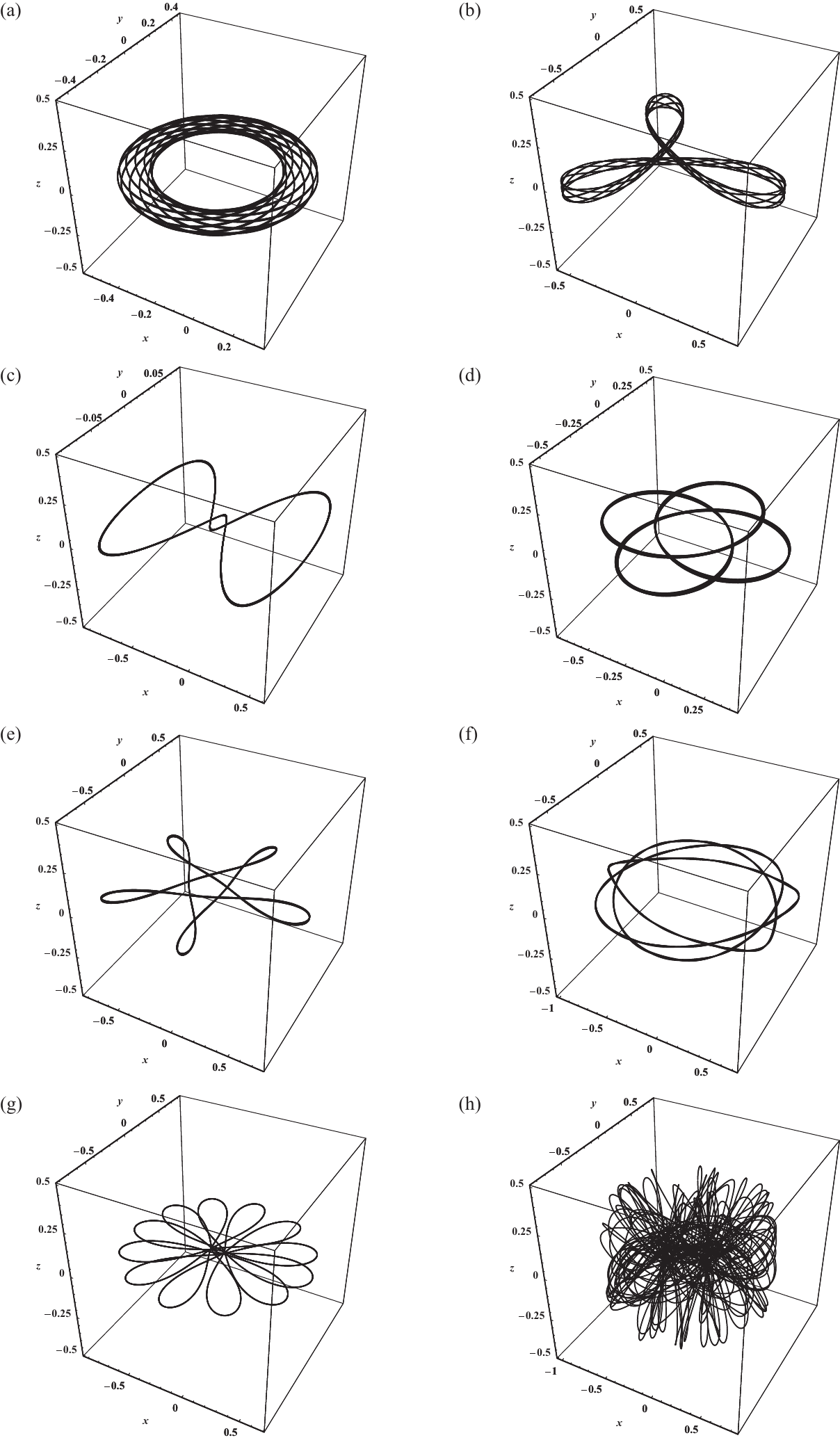}}}
\vskip 0.01cm
\caption{\small (a--h) Eight representative regular orbits of the 3D dynamical system. The initial conditions and more details regarding the values of all the other parameters are given in the text.}
\end{figure*}

In Figures 16(a--h) we present eight typical orbits of the 3D dynamical system. Figure 16(a) shows a regular quasi-periodic orbit circulating around the primary galaxy, with initial conditions $x_0=0.167$, $y_0=0$, $z_0=0.01$, and $p_{x0}=p_{z0}=0$, while the value of $p_{y0}$ is always found from the Jacobi integral (10). In Figure 16(b), we observe a 3D periodic orbit characteristic of the 1:2 resonance. This orbit has initial conditions $x_0=-0.1895$, $y_0=0$, $z_0=0.02$, and $p_{x0}=p_{z0}=0$. Figure 16(c) shows a 3D periodic orbit with initial conditions $x_0=0.5788$, $y_0=0$, $z_0=0.01$, and $p_{x0}=p_{z0}=0$, circulating around the center of the primary galaxy. In Figure 16(d), we see a quasi-periodic orbit, with initial conditions $x_0=0.4295$, $y_0=0$, $z_0=0.015$, and $p_{x0}=p_{z0}=0$. A 3D resonant orbit with initial conditions $x_0=0.7445$, $y_0=0$, $z_0=0.01$, and $p_{x0}=p_{z0}=0$, moving around the galactic center, is shown in Figure 16(e). In Figure 16(f) we observe a 3D periodic orbit circulating around the primary galaxy. This orbit has initial conditions $x_0=-0.9854, y_0=0, z_0=0.01, \text{ and }p_{x0}=p_{z0}=0$. In Figure 16(g) we observe a complicated 3D resonant periodic orbit with initial conditions $x_0=0.673, y_0=0, z_0=0.005, \text{ and }p_{x0}=p_{z0}=0$. Figure 16(h) shows a 3D chaotic orbit with initial conditions $x_0=-0.31$, $y_0=0$, $z_0=0.1$, and $p_{x0}=p_{z0}=0$. This orbit goes arbitrarily close to the primary galaxy and it is deflected to higher values of $z$, on approaching the dense and massive nucleus. We must note that in all 3D orbits shown in Figures 16(a--h), the initial conditions $\left(x_0, p_{x0}\right)$ and the values of all the other parameters $\left(M_{\rm d}, M_{\rm n}, R\right)$ are as in the corresponding 2D orbits shown in Figures 13(a--h). Moreover, we observe that all regular 3D orbits shown in Figures 16(a--g) stay near the galactic plane and, therefore, support the disk structure of the primary galaxy. The numerical integration time for all 3D orbits shown in Figures 16(a--h) is 200 time units.

\section{A Theoretical Approach}

In this section, we shall present some theoretical arguments, together with elementary numerical calculations, in order to explain the numerically obtained relationships given in Figures 11, 12, and 14. Moreover, we will quote different kinds of theoretical techniques in an attempt to explain and study deeper the dynamical structure of the 2D or 3D Hamiltonian system.

Since the potential of the nucleus is integrable with spherical symmetry, the observed chaotic phenomena in our dynamical system should derive mainly from the total $| F_t |$ force. Figure 17 shows a plot of the $| F_t |$ force as a function of the distance $R$ between the two galaxies at the point $P_c(x_0,y_0,z_0)=(-0.12, 0.02, 0.01)$. This point is very close to the center of the primary galaxy, where the chaotic phenomena are more prominent. One may observe, from the plot shown in Figure 20 that the two curves, which correspond to two different cases (active and quiet primary galaxy), have almost the same pattern. In both cases, the $| F_t |$ force decreases, tending asymptotically to zero, as the value of the distance $R$ increases. Here, we must point out that the values of the $| F_t |$ force are larger when the primary galaxy is active and, therefore, the corresponding curve is above the one corresponding to the quiet galaxy. Thus, we conclude that when a massive and dense nucleus is present in the center of the primary galaxy, the extent of the chaotic orbits is larger than in the case in which the primary galaxy is quiet. One can see that the pattern between the $| F_t |$ force and the distance $R$ shown in Figure 17 and obtained theoretically is very similar to those connecting the chaotic percentage $A\%$ or the LCE with the distance $R$ shown in Figures 11, 12, and 14, respectively, which have been obtained numerically. Thus, the numerically obtained results given in Figures 11, 12, and 14 regarding the evolution of the chaos with respect to the distance $R$ can be justified and explained, in a way, theoretically through the relation between the strength of the total force and the distance $R$ depicted in Figure 17.
\begin{figure}[!tH]
\centering
\resizebox{\hsize}{!}{\rotatebox{0}{\includegraphics*{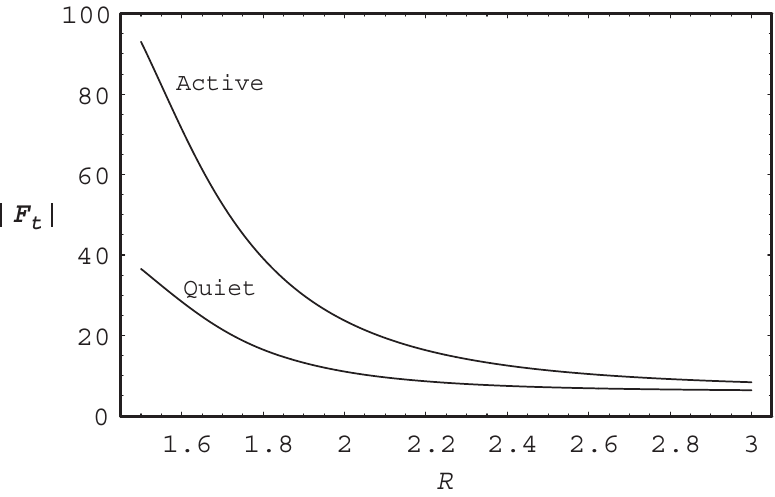}}}
\caption{\small A plot of the total $| F_t |$ force as a function of the distance $R$, when the primary galaxy is active or quiet.}
\end{figure}

Furthermore, it would be of particular interest to study the structure of the velocity profile that is the plot of the total velocity of a test particle (star), $\upsilon = \sqrt{\dot{x}^2 + \dot{y}^2 + \dot{z}^2}$ as a function of time, for ordered and chaotic motion. Figure 18(a) shows the velocity profile for a time period of 500 time units corresponding to the regular 3D orbit shown in Figure 16(a). Here, the velocity profile is quasi-periodic and the maximum value of the velocity is about 450 km\,s\(^{-1}\). This suggests that the regular 3D motion occurs in small velocities. Figure 18(b) shows the velocity profile for a time period of 500 time units corresponding to the chaotic 3D orbit shown in Figure 16(h). In this case, we can discuss two aspects. The first is that the velocity obtains high values up to 790 km\,s\(^{-1}\) and the second is that the velocity profile appears to be highly asymmetric, displaying abrupt changes and large deviations between the maxima and also between the minima in the $\left[\upsilon - t \right]$ plot. Therefore, we may conclude that the chaotic 3D motion occurs in high and abruptly changing velocities.
\begin{figure*}[!tH]
\centering
\resizebox{\hsize}{!}{\rotatebox{0}{\includegraphics*{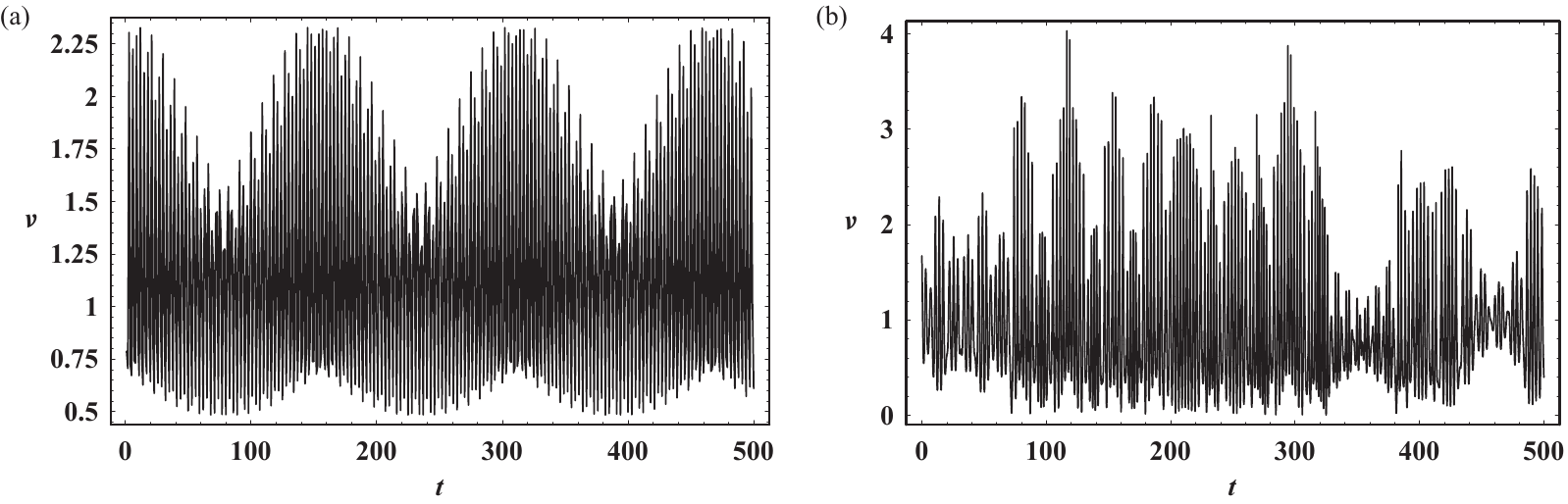}}}
\vskip 0.01cm
\caption{\small (a) The total velocity profile of the 3D orbit shown in Figure 16(a). We observe a nearly periodic pattern. (b) The total velocity profile for the chaotic 3D orbit shown in Figure 16(h). In this case, there are abrupt changes in the profile's pattern indicating chaotic motion.}
\end{figure*}
\begin{figure*}[!tH]
\centering
\resizebox{\hsize}{!}{\rotatebox{0}{\includegraphics*{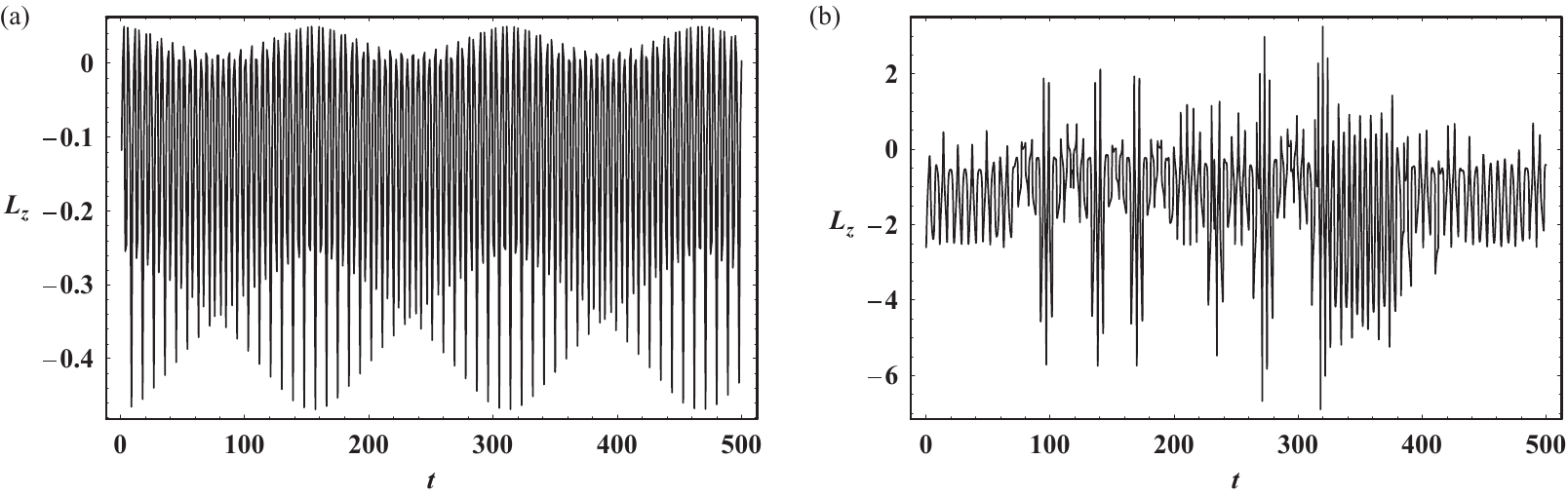}}}
\vskip 0.01cm
\caption{\small A plot of the $L_z$ component of the total angular momentum versus time for (a) the regular 3D orbit shown in Figure 16(a) and (b) the chaotic 3D orbit shown in Figure 16(h).}
\end{figure*}

The physical parameter playing an important role in the orbital behavior of the stars is the $L_z$ component of the total angular momentum. From our previous experience, we know that  on approaching a dense and massive nucleus, low angular momentum stars are scattered off the galactic plane, displaying chaotic motion (Caranicolas \& Innanen 1991; Caranicolas \& Papadopoulos 2003; Caranicolas \& Zotos 2010; Zotos 2011a, 2011c). Of course, here in 3D phase space, things are more complicated than in axially symmetric dynamical models, where the $L_z$ component is conserved. As the motion takes place in a rotating non-axially symmetric system, the $L_z$ component is not conserved and it is given by
\begin{equation}
L_z = x \dot{y} - \dot{x} y - \Omega_p \left(x^2 + y^2\right).
\end{equation}
Nevertheless, we can compute numerically its mean value $\langle  L_z\rangle$ using the formula
\begin{equation}
\langle L_z\rangle = \frac{1}{n} \displaystyle\sum\limits_{i=0}^n L_{zi}.
\end{equation}
Our numerical calculations suggest that the chaotic orbits have low values of $\langle L_z\rangle$, while regular orbits obtain high values of $\langle L_z\rangle$. Figure 19(a) shows a plot of the evolution of the $L_z$ component with the time for the regular orbit of Figure 16(a). In this case, we observe that $L_z$ is nearly a quasi-periodic function of time, while $\langle L_z\rangle=-0.12271$. Figure 19(b) is similar to Figure 19(a) but for the chaotic orbit shown in Figure 16(h). Here one can see abrupt changes in $L_z$ during the chaotic motion, while for this chaotic orbit we have $\langle L_z \rangle = -1.24662$. In both cases, the time interval of the numerical integration is 500 time units, while $n =10^4$.
\begin{figure}[!tH]
\centering
\resizebox{\hsize}{!}{\rotatebox{0}{\includegraphics*{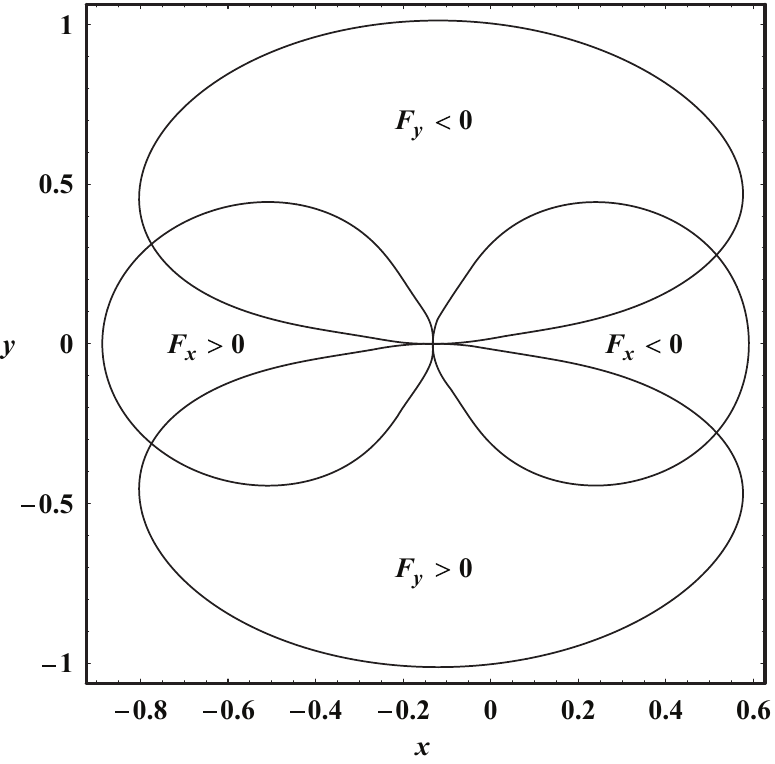}}}
\caption{\small Contours of the $F_x=const$ (elliptic shaped) together with the contours $F_y=const$ (figure-eight-shaped). Details are given in the text.}
\end{figure}
\begin{figure*}[!tH]
\centering
\resizebox{\hsize}{!}{\rotatebox{0}{\includegraphics*{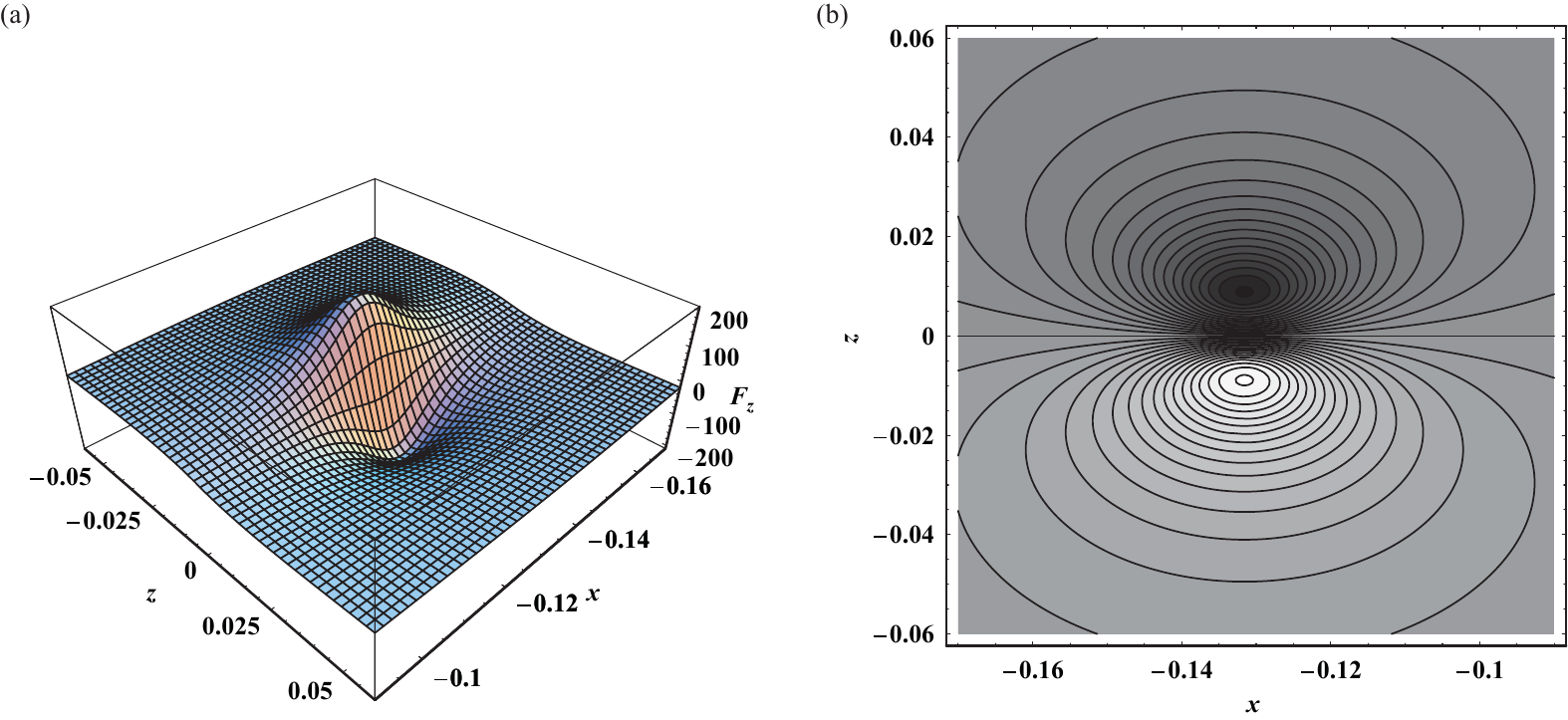}}}
\vskip 0.01cm
\caption{\small (a) A 3D plot of the value of the $F_z$ force on the $(x,z)$ plane and (b) contours of the projections $F_z=const$ on the $(x, z)$ plane. Lighter colors indicate higher values of $F_z$. For positive values of $z$ the $F_z$ force is negative, while for negative values of $z$ the $F_z$ force is positive.}
\end{figure*}
\begin{figure}[!tH]
\centering
\resizebox{\hsize}{!}{\rotatebox{0}{\includegraphics*{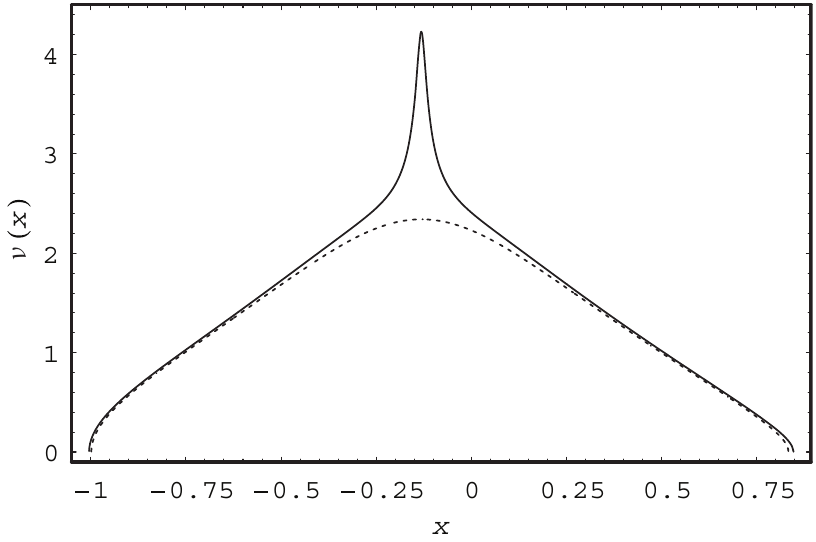}}}
\caption{\small A plot of the total velocity $\upsilon (x)$ as a function of the distance $x$, for the two cases (active and quiet primary galaxy).}
\end{figure}

In what follows we shall present a semi-theoretical analysis, in order to give a more detailed picture of the structure of the dynamical system and its behavior. The forces acting on a test particle along the $x$ and $y$ axes are given by the equations
\begin{eqnarray}
F_x &=& -\frac{M_{\rm n}\left(x - x_1\right)}{\left(r_{a1}^2 + c_{\rm n}^2\right)^{3/2}} -
\frac{M_{\rm d}\left(x - x_1\right)}{\left[b^2 + r_{a1}^2 + \left(a + h\right)^2\right]^{3/2}} \nonumber\\
&-& \frac{M_{\rm s}\left(x - x_2\right)}{r_{a2}^3} + \Omega_p^2 x\ - 2\Omega_p \dot{y}, \nonumber\\
F_y &=& -\frac{M_{\rm n} y}{\left(r_{a1}^2 + c_{\rm n}^2\right)^{3/2}} -
\frac{M_{\rm d} y}{\left[b^2 + r_{a1}^2 + \left(a + h\right)^2\right]^{3/2}} \nonumber\\
&-& \frac{M_{\rm s} y}{r_{a2}^3} + \Omega_p^2 y + 2\Omega_p \dot{x}.
\end{eqnarray}
It is obvious from Equations (17) that the strength of both forces increases as the mass of the nucleus or the disk increases or their scale lengths decrease. Figure 20 shows the contours of $F_x=const$ together with the contours $F_y=const$. The values of all other parameters are as in Figure 2. The contours of $F_x=const$ look like ellipses, while the contours of $F_y=const$ have a figure-eight shape. Inside the first `ellipse' we have $F_x > 0$, while inside the second `ellipse' we have $F_x <  0$. On the other hand, inside the lower figure-eight curve we have $F_y > 0$, while inside the upper figure-eight curve we have $F_y < 0$. Looking carefully near the galactic center, we observe that there are areas where $F_x$ is positive and $F_y$ is negative at the same time and vice versa. There are also areas where both forces are positive or negative at the same time. Therefore, we conclude that near the center of the primary galaxy, there are strong attractive and repulsive forces acting on the test particle (star). These forces are responsible for the chaotic scattering of the star near the galactic center, leading to chaotic motion.

At this point, we can expand the semi-theoretical analysis described above, in order to give a more detailed and complete picture of the structure of the 3D dynamical system and its behavior. The force acting on a test particle along the $z$ axis is given by the equation
\begin{equation}
F_z = - \frac{M_{\rm n} z}{\left(r_1^2 + c_{\rm n}^2\right)^{3/2}}
- \frac{M_{\rm d} \left(a + r_z\right)z}{r_z \left[b^2 + r_{a1}^2 + \left(a + r_z\right)^2\right]^{3/2}}
- \frac{M_{\rm s} z}{r_2^3},
\end{equation}
where $r_z = \sqrt{h^2 + z^2}$. It is obvious from Equation (18) that the strength of the $F_z$ force increases as the mass of the nucleus or the disk increases or their scale lengths decrease. Figure 21(a) shows a 3D plot of the value of the $F_z$ force on the $(x, z)$ plane. Figure 21(b) depicts the contours of the projections $F_z = const$ on the $(x, z)$ plane. We observe that for positive values of $z,$ the $F_z$ force is negative, while for negative values of $z,$ the value of the $F_z$ is positive, near the center of the primary galaxy. Therefore, lighter colors on the lower half part of the $(x, z)$ plane indicate higher values of the $F_z$ force, while darker colors on the upper half part of the same plane indicate lower values of the $F_z$ force. One can observe that near the active galactic core which hosts a dense and massive nucleus, the test particle experiences a very strong $F_z$ force. The values of all the parameters in Figures 21(a--b) are as in Figure 2.

Last but not least, we also investigate the behavior of the total velocity near the galactic core as a function of the distance $x$. To do that, we consider the limiting curve, that is, the curve containing all the invariant curves on the $(x, p_x)$ phase plane. This can be obtained if we set $y=z=p_y=p_z=0$ in Equation (10), yielding
\begin{equation}
\frac{1}{2}p_x^2 = \frac{1}{2}\upsilon^2 = E_J - \Phi_{\rm t} (x),
\end{equation}
where we have set $p_x=\upsilon$ because at the limiting curve the $p_x$ velocity is the total velocity (see also Papadopoulos \& Caranicolas 2006). In Figure 22, we observe two plots of the total velocity $\upsilon$ as a function of the distance $x$, derived using relation (19). The values of all the parameters are as in Figure 2. The dashed line corresponds to the case of the quiet primary galaxy, while the solid line corresponds to the case when the primary galaxy hosts a nucleus in its core and, therefore, is active. We see that the velocity is about the same for all values of $x$, except near the galactic center, that is when $x \simeq -0.13$, where higher velocities correspond to the active galactic center, hosting massive and dense nucleus.
\begin{table}[ht]
\setlength{\tabcolsep}{4.0pt}
\centering
\caption{Radii of Lindblad Resonances When $R=2.35$ and $\Omega_p = 0.419146$, for Both Active and Quiet Primary Galaxies}
\ \ \ \\
\begin{tabular}{|c|c|c||c|c|}
\hline
{Case} & {Resonance} & {Region} & $r_1$ & $r_2$ \bigstrut[t] \\
\hline \hline
\multirow{8}[8]{*}{Active} &
\multirow{2}[4]{*}{$\Omega_p = \Omega - 2 \kappa /3$}
          & Direct
                       & 0.0149 &   --    \bigstrut[t] \\ \cline{3-5}
&         & Retrograde & 0.0142 & 0.2385 \bigstrut[t] \\ \cline{2-5} &
\multirow{2}[4]{*}{$\Omega_p = \Omega - 3 \kappa /4$}
          & Direct
                       & 0.0246 &   --    \bigstrut[t] \\ \cline{3-5}
&         & Retrograde & 0.0217 & 0.3548 \bigstrut[t] \\ \cline{2-5} &
\multirow{2}[4]{*}{$\Omega_p = \Omega - 5 \kappa /8$}
          & Direct
                       & 0.0127 & 0.3886 \bigstrut[t] \\ \cline{3-5}
&         & Retrograde & 0.0122 &   --    \bigstrut[t] \\ \cline{2-5} &
\multirow{2}[4]{*}{$\Omega_p = \Omega - 7 \kappa /9$}
          & Direct
                       & 0.0257 &   --    \bigstrut[t] \\ \cline{3-5}
&         & Retrograde & 0.0238 & 0.3912 \bigstrut[t] \\
\hline
\multirow{8}[8]{*}{Quiet} &
\multirow{2}[4]{*}{$\Omega_p = \Omega - 2 \kappa /5$}
          & Direct
                       & 0.0123 & 0.5527 \bigstrut[t] \\ \cline{3-5}
&         & Retrograde & 0.0114 & 0.1342 \bigstrut[t] \\ \cline{2-5} &
\multirow{2}[4]{*}{$\Omega_p = \Omega - 3 \kappa /4$}
          & Direct
                       & 0.0120 &   --    \bigstrut[t] \\ \cline{3-5}
&         & Retrograde & 0.0119 & 0.2351 \bigstrut[t] \\ \cline{2-5} &
\multirow{2}[4]{*}{$\Omega_p = \Omega - 3 \kappa /7$}
          & Direct
                       & 0.0095 & 0.6782 \bigstrut[t] \\ \cline{3-5}
&         & Retrograde & 0.0091 &   --    \bigstrut[t] \\ \cline{2-5} &
\multirow{2}[4]{*}{$\Omega_p = \Omega - 4 \kappa /9$}
          & Direct
                       & 0.0207 &   --    \bigstrut[t] \\ \cline{3-5}
&         & Retrograde & 0.0196 & 0.2608 \bigstrut[t] \\
\hline
\end{tabular}
\end{table}

One of the factors responsible for the chaotic motion and other resonance phenomena, such as islandic motion corresponding to sets of multiple islands of invariant curves or invariant tori, are the several inner Lindblad resonances
\begin{equation}
\Omega_p = \Omega - \frac{n}{m} \kappa,
\end{equation}
where $\Omega$ and $\kappa$ indicate the circular and the epicycle frequency of the star, respectively, while $m$ and $n$ are integers. The main resonances for the two cases (active and quiet primary galaxy) together with the corresponding resonance radii $r_1$ and $r_2$, when $R=2.35$ and $\Omega_p = 0.419146,$ are given in Table 2.
\begin{figure*}[!tH]
\centering
\resizebox{\hsize}{!}{\rotatebox{0}{\includegraphics*{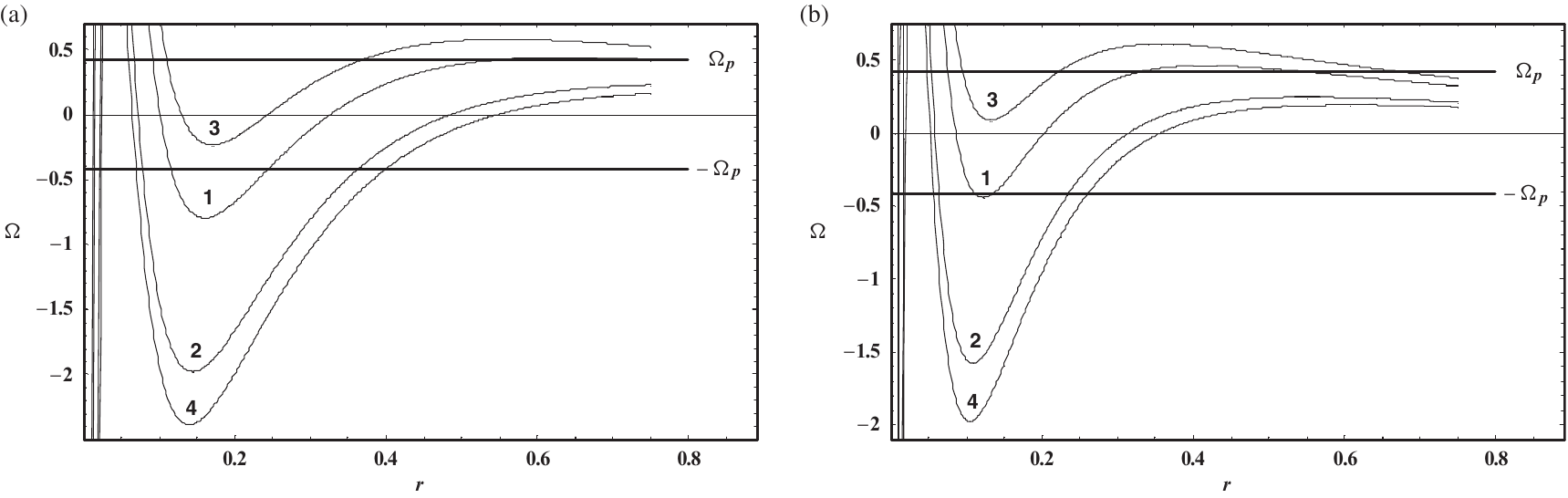}}}
\vskip 0.01cm
\caption{\small The curves $\Omega-n\kappa/m$ versus $r$, when $R=2.35$ and $\Omega_p = 0.419146$, when (a) the primary galaxy is active and (b) the primary galaxy is quiet.}
\end{figure*}

Figure 23(a) shows a plot of the curves $\Omega-n\kappa/m$ in the case where the primary galaxy is active, as a function of the radius $r$, when $R=2.35$. We choose this particular value of the distance between the two galaxies because at this distance we can observe a large variety of Lindblad resonances. The numbers 1, 2, 3, and 4 indicate the curves $\Omega-2\kappa/3$, $\Omega-3\kappa/4$, $\Omega-5\kappa/8,$ and $\Omega-7\kappa/9,$ respectively. The straight lines are the curves $\Omega_p=\pm 0.419146$. The values of all the other parameters are as in Figure 2. Figure 23(b) is similar to Figure 23(a) but when the primary galaxy is quiet. Here, the numbers 1, 2, 3, and 4 indicate the curves $\Omega-2\kappa/5$, $\Omega-3\kappa/4$, $\Omega-3\kappa/7,$ and $\Omega-4\kappa/9,$ respectively. As we can see, there are a considerable number of resonance radii for both the direct and retrograde orbits in both cases. Details regarding the resonances and the resonance radii can be obtained from Table 2. In other words, all the Lindblad resonances given in Table 2 are also responsible for the chaotic motion in the primary galaxy. It is also interesting to note that the above resonances produce large chaotic regions for small values of the distance $R$, while for larger values of $R$ (see Figures 9a--b) the chaotic regions are extremely small, although the resonance radii are still present. This means that in this case the distance between the two galaxies precedes the Lindblad resonances.
\begin{figure*}[!tH]
\centering
\resizebox{\hsize}{!}{\rotatebox{0}{\includegraphics*{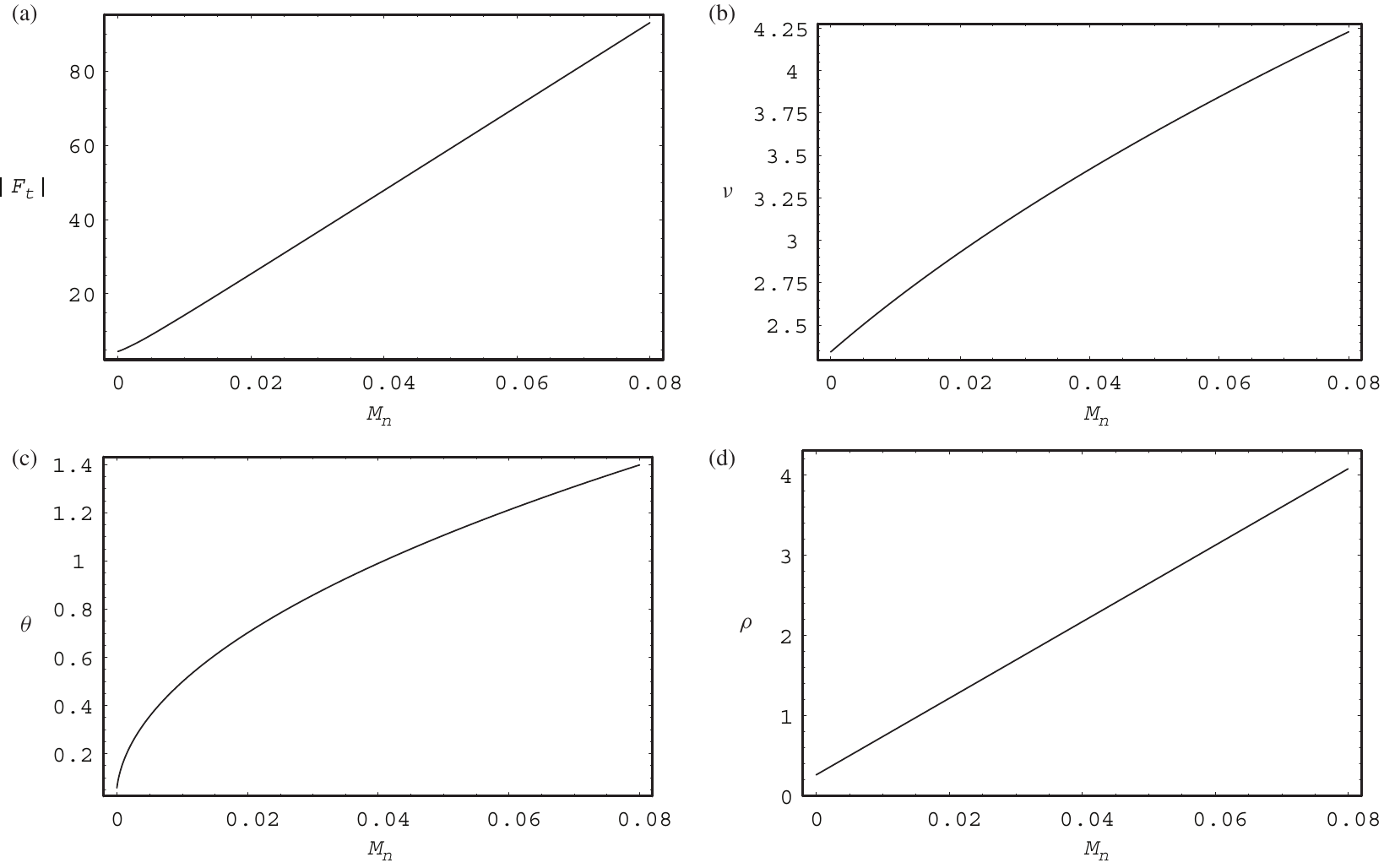}}}
\vskip 0.01cm
\caption{\small (a--d) Evolution of different dynamical quantities of the system as a function of the mass of the nucleus $M_{\rm n}$. Details are given in the text.}
\end{figure*}

In all cases, the primary galaxy is assumed to have a fixed value of total mass
\begin{equation}
M_\text{P} = M_{\rm d} + M_{\rm n} = 2.08,
\end{equation}
where $0 \leq M_{\rm n} \leq 0.08$. For a constant distance $R,$ the $F_x$, $F_y,$ and $F_z$ forces, per unit mass, at an arbitrary but fixed point $P_0(x_0, y_0, z_0)$ near the center of the primary galaxy are linear functions of the mass of the nucleus $M_{\rm n}$, of the form
\begin{eqnarray}
F_x &=& -c_1 -k_1 M_{\rm n}, \nonumber \\
F_y &=& -c_2 -k_2 M_{\rm n}, \nonumber \\
F_z &=& -c_3 -k_3 M_{\rm n},
\end{eqnarray}
where $(c_1, c_2, c_3)$ and $(k_1, k_2, k_3)$ are positive constants, while relation (21) has also been taken into account. Figure 24(a) shows the total $| F_t |$ force as a function of $M_{\rm n}$, at $P_0(x_0, y_0, z_0)= (-0.12, 0.02, 0.01)$, when $R=1.5$. We observe that the total force increases rapidly as the mass of the nucleus $M_{\rm n}$ increases. The presence of the nucleus increases the velocity near the center of the primary galaxy. This can be easily shown if we consider the limiting curve, that is, the curve enclosing all the invariant curves in the $(x,p_x)$ phase plane. This is obtained if we set $y=0$ in the Jacobi integral (13). Thus, we have
\begin{equation}
p_y^2 = 2\left[E_J - \Phi_{\rm t}(x) \right] - p_x^2 \geq 0.
\end{equation}
The curve
\begin{equation}
f\left(x,p_x\right) = 2\left[E_J - \Phi_{\rm t}(x) \right] - p_x^2 = 0
\end{equation}
is a curve on the $(x, p_x)$ phase plane, called the limiting curve or the zero velocity curve (ZVC). The velocity $\upsilon_x = p_x$ at the limiting curve is the total velocity $\upsilon$. At a fixed point $x = x_0,$ the total velocity is
\begin{equation}
\upsilon = \left[2 E_J + const + c_4 \left(x_0 - x_1 \right)^2
+ \frac{2 M_{\rm n} \left(\lambda_1 - \lambda_2 \right)}{\lambda_1 \lambda_2} \right]^{1/2},
\end{equation}
where $c_4 < 1$ is a constant, while
\begin{eqnarray}
\lambda_1 &=& \sqrt{b^2 + \left(x_0 - x_1 \right)^2 + (a + h)^2}, \nonumber \\
\lambda_2 &=& \sqrt{\left(x_0 - x_1 \right)^2 + c_{\rm n}^2}.
\end{eqnarray}
As $\lambda_1 > \lambda_2$, the maximum value of the velocity is met when $x_0 = x_1$ (see also Figures 7a--b). It is evident from Equations (25) and (26) that $\upsilon$ increases as $M_{\rm n}$ increases or $c_{\rm n}$ decreases. This indicates that high velocities are expected in the central regions of galaxies with dense and massive nuclei. Figure 24(b) shows a plot of $\upsilon$ as a function of $M_{\rm n}$ when $R=1.5$. We see that the velocity increases as the mass of the nucleus $M_{\rm n}$ increases. In Figure 24(c) we observe a plot of the rotational velocity $\Theta$ of the primary galaxy, as a function of the mass of the nucleus $M_{\rm n}$, when $r=r_0=0.01$. The exact function is
\begin{equation}
\Theta (M_{\rm n}) = r_0 \sqrt{\frac{M_{\rm n}}{\left(r_0^2 + c_{\rm n}^2\right)^{3/2}} +
\frac{M_{\rm d}}{\left(b^2 + r_0^2 + \left(a + h \right)^2\right)^{3/2}}}.
\end{equation}
From Figure 24(c) we can derive two basic conclusions. (i) In galaxies with massive and dense central concentrations, the rotational velocity has and retains high values near the galactic center. (ii) The more massive and dense the nucleus, the higher is the value of the rotational velocity $\Theta$. Figure 24(d) shows the evolution of the mass density $\rho$ of the primary galaxy, as a function of the mass of the nucleus $M_{\rm n}$ at the point $P_0(x_0,y_0,z_0)=(-0.12,0.02,0.01)$. It is evident that as the galaxy evolves and a nucleus forms in its center, we have a constant and rapid increase of the mass density. The relationships given in Figures 24(a--d) are of great importance and will prove very useful in the following section, where we will study the evolution of our galactic system using a time-dependent model.
\begin{figure*}[!tH]
\centering
\resizebox{\hsize}{!}{\rotatebox{0}{\includegraphics*{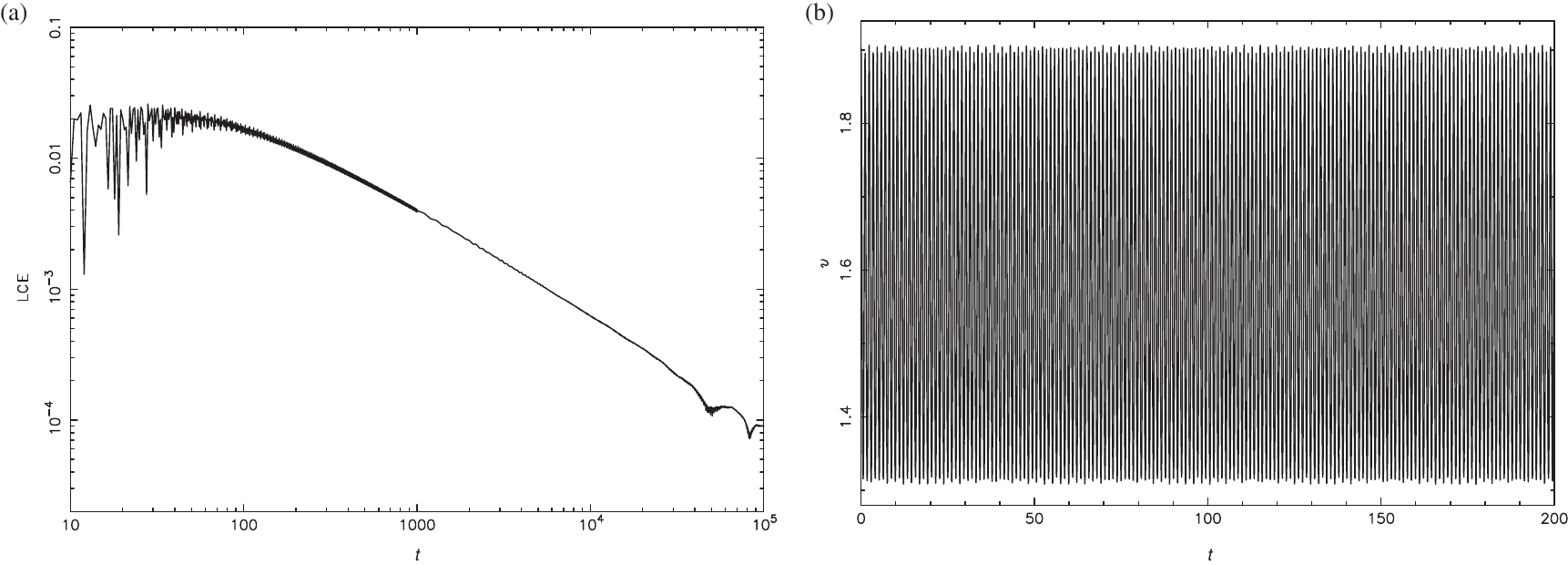}}}
\vskip 0.01cm
\caption{\small (a) Evolution of the LCE of the 3D orbit in the time-dependent model, following relations (28), and (b) the corresponding velocity profile. The orbit starts as a regular and remains regular during the mass transportation.}
\end{figure*}
\begin{figure*}[!tH]
\centering
\resizebox{\hsize}{!}{\rotatebox{0}{\includegraphics*{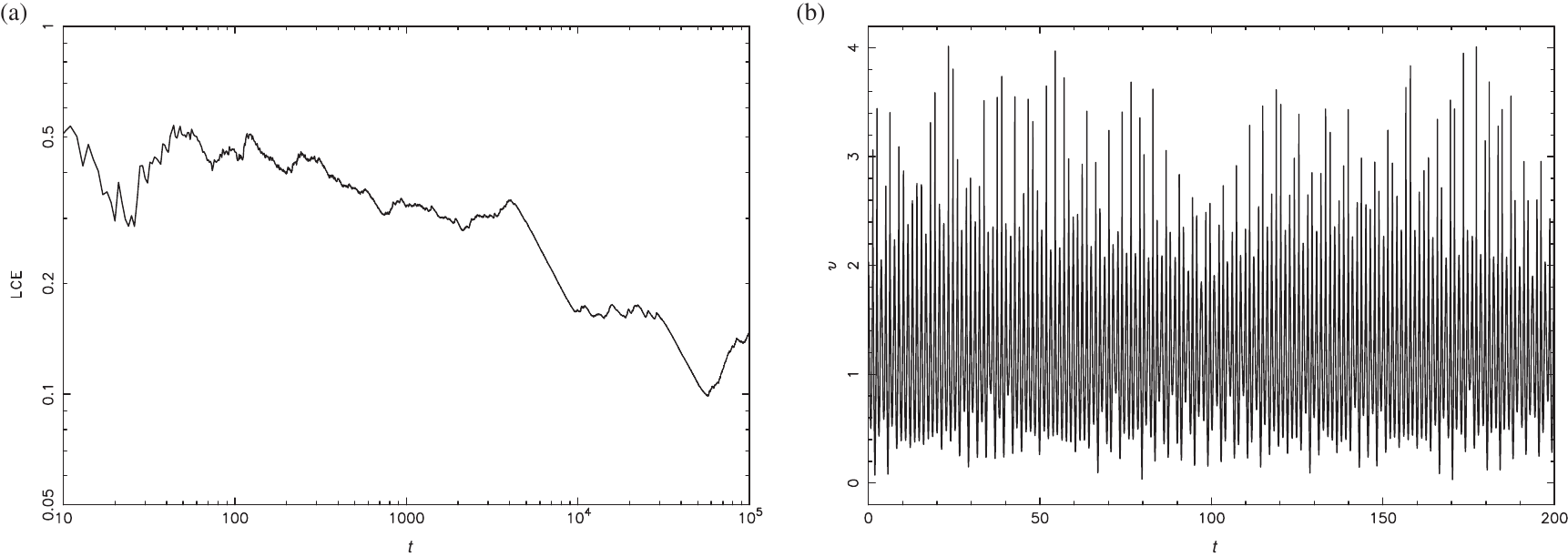}}}
\vskip 0.01cm
\caption{\small (a) Evolution of the LCE of the 3D orbit in the time-dependent model, following relations (28), and (b) the corresponding velocity profile. The orbit starts as a chaotic and remains chaotic during the mass transportation.}
\end{figure*}
\begin{figure*}[!tH]
\centering
\resizebox{\hsize}{!}{\rotatebox{0}{\includegraphics*{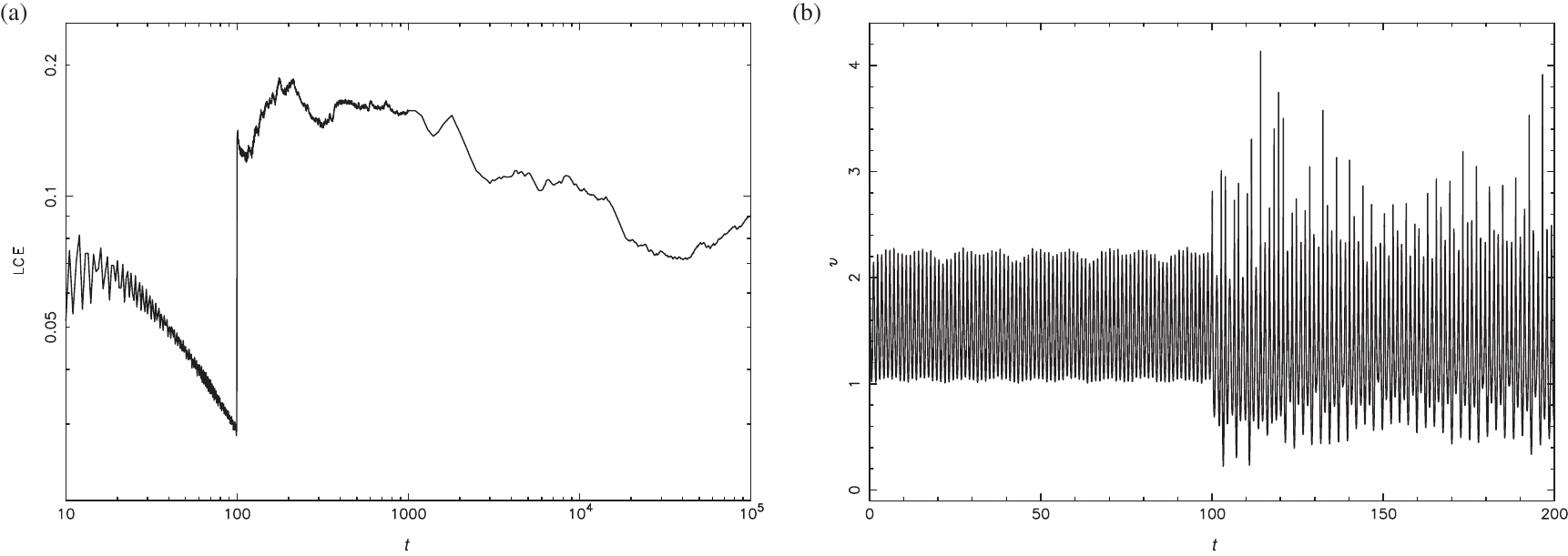}}}
\vskip 0.01cm
\caption{\small (a) Evolution of the LCE of the 3D orbit in the time-dependent model, following relations (28), and (b) the corresponding velocity profile. The orbit starts as a regular but after 100 time units; when the galactic evolution stops it becomes chaotic.}
\end{figure*}
\begin{figure*}[!tH]
\centering
\resizebox{\hsize}{!}{\rotatebox{0}{\includegraphics*{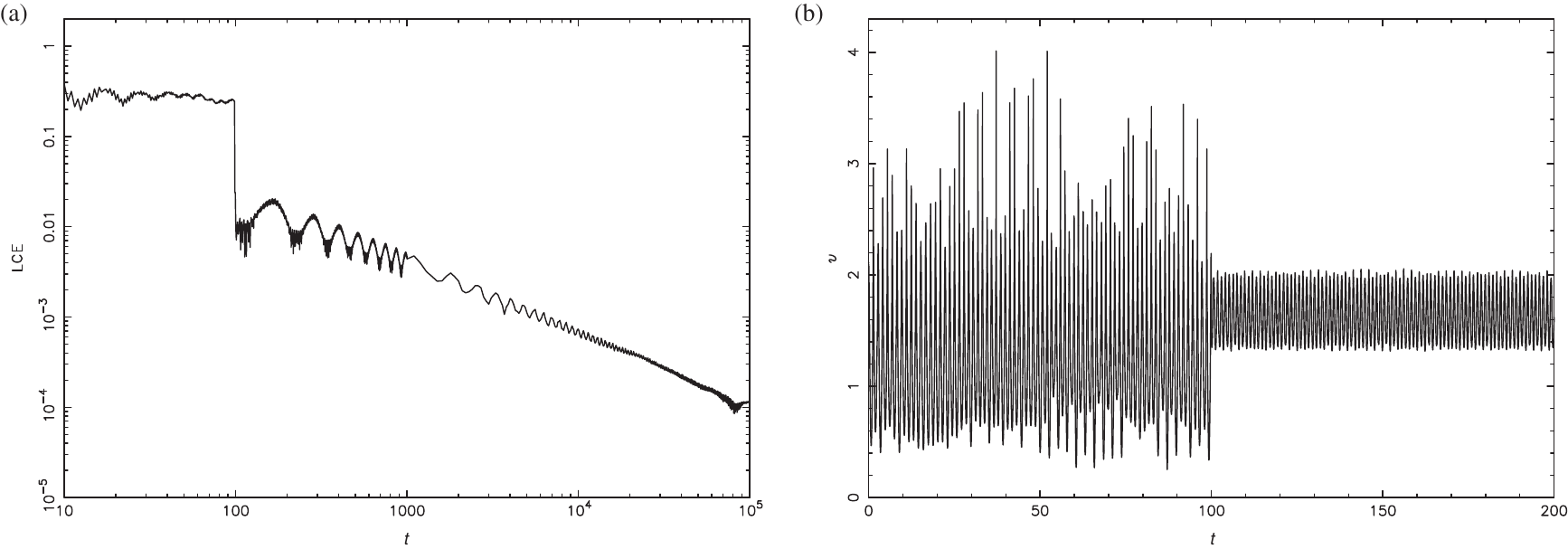}}}
\vskip 0.01cm
\caption{\small (a) Evolution of the LCE of the 3D orbit in the time-dependent model, following relations (28), and (b) the corresponding velocity profile. The orbit starts as a chaotic but after the galactic evolution, it changes its nature to regular.}
\end{figure*}

\section{Evolution of the Orbits in the 3D Time-Dependent Model}

Let us now follow the evolution of 3D orbits, as mass is transported from the disk of the quiet galaxy to its center. By this procedure, a massive and dense nucleus is developed in the central regions of the primary galaxy. The mass transport is linear following the set of equations:
\begin{eqnarray}
M_{\rm nf}(t) &=& M_{\rm ni} + k t, \nonumber \\
M_{\rm df}(t) &=& M_{\rm di} - k t,
\end{eqnarray}
where $M_{\rm ni}=0$ and $M_{\rm di}=2.08$ are the initial values of the mass of the nucleus and the disk respectively, while $k$ is a positive parameter. We assume that the linear rate described by relations (28) is slow compared to the orbital period of the binary system and therefore it is adiabatic. This is true because the mass transportation period is 100 time units, while the orbital period is about three orders of magnitude smaller. It is also assumed that the transportation stops when the mass of the nucleus in the galactic core takes the value $M_{\rm nf}=0.08$. It is well known that the shape of the 3D orbits sometimes is inconclusive or even misleading. In order to overcome this drawback, we have decided to use a more accurate method such as the LCE. The main advantage of this method is that it uses certain and objective numerical thresholds, beyond which we can distinguish between ordered and chaotic motion. On the other hand, this method is very time-consuming as it needs time intervals of numerical integration of the order of $10^5$ time units, in order to provide reliable and definitive results regarding the nature of a 3D orbit. But this is a `price' we can afford to pay. Additionally, we shall use a fast but qualitative method, that is the profile's pattern of the total velocity, so as to characterize a 3D orbit. In the following, we present the evolution of four different 3D orbits, as the total mass distribution of the dynamical system changes with time, following the set of equations (28). For all orbits shown in Figures 25--28, the initial value of the Hamiltonian (10), at $t=0$, is $E_{Ji}=-2.6$, while $k=0.0008$.

In Figure 25(a) we can see the evolution of the LCE for a 3D orbit and for a time period of $10^5$ time units, as the galaxy evolves following the set of equations (28). The initial conditions are $x_0=0.12$, $y_0=0$, $z_0=0.02$, $p_{x0}=0$, and $p_{z0}=0$, while the value of $p_{y0}$ is found from the Hamiltonian (10) in all cases. The values of all the other parameters are as in Figure 7(b). When $t=100$ time units, the mass of the developed nucleus in the core of the galaxy is $M_{\rm nf}=0.08$ and the evolution stops. The value of the Hamiltonian is now $E_{Jf}=-2.60012$. The profile of the LCE clearly indicates that this orbit starts as a regular and remains regular during the galactic evolution. Figure 25(b) shows the velocity profile for the same 3D orbit for a time period of 200 time units. The pattern is almost quasi-periodic and therefore the orbit remains regular during the galactic evolution.

On the other hand, things are quite different in Figure 26(a). This orbit has initial conditions $x_0=-0.44$, $y_0=0$, $z_0=0.01$, $p_{x0}=0$, and $p_{z0}=0$. The values of all the other parameters are as in Figure 7(b). When $t=100$ time units, the mass of the developed nucleus in the core of the primary galaxy is $M_{\rm nf}=0.08$ and the mass transportation stops. The value of the Hamiltonian is now $E_{Jf}=-2.60057$. In this case, the profile of the LCE clearly indicates that this orbit starts as a chaotic 3D orbit and remains chaotic during the galactic evolution. Figure 26(b) shows the velocity profile for the same 3D orbit, for a time period of 200 time units. The pattern is highly asymmetric, with a large number of abrupt peaks, large deviations between the maxima and also significant deviations between the minima. Therefore, we conclude that this orbit remains chaotic during the mass transportation.

In Figure 27(a) we observe the evolution of the LCE of a 3D orbit with initial conditions $x_0=-0.196$, $y_0=0$, $z_0=0.01$, $p_{x0}=0$, and $p_{z0}=0$. The values of all the other parameters are as in Figure 7(b). After a time interval of 100 time units the mass of the nucleus in the core of the galaxy is $M_{\rm nf}=0.08$ and the galactic evolution stops. The Hamiltonian settled to the value $E_{Jf}=-2.60034$. The profile of the LCE given in Figure 27(a) shows that the orbit starts as a regular 3D orbit, but after the galactic evolution it becomes chaotic. It is evident that if mass transport were not present, the orbit would have remained regular. The presence of the nucleus in the core of the primary galaxy has changed the character of the 3D orbit from regular to chaotic. The same result can be obtained by looking at the profile of the velocity of the 3D orbit, as shown in Figure 27(b). For the first 100 time units the pattern is quasi-periodic, indicating regular motion, but after that time interval it becomes highly asymmetric, leading to the conclusion that the orbit has become chaotic.

Figure 28(a) depicts the evolution of the LCE of a 3D orbit with initial conditions $x_0=0.74$, $y_0=0$, $z_0=0.01$, $p_{x0}=0$, and $p_{z0}=0$. The values of all the other parameters are as in Figure 7(b). After a time interval of 100 time units, the mass of the nucleus in the core of the galaxy is $M_{\rm nf}=0.08$ and the mass transportation stops. The Hamiltonian settled to the value $E_{Jf}=-2.60021$. The profile of the LCE given in Figure 28(a) indicates that this orbit starts as a chaotic 3D orbit, but after the galactic evolution it becomes an ordered one. Therefore, it is evident that if mass transportation were not present, the orbit this time would have remained chaotic. In this case, the presence of the nucleus in the core of the galaxy has changed the nature of the 3D orbit from chaotic to regular. A similar result can be obtained from the profile of the velocity of the 3D orbit, as shown in Figure 28(b). For the first 100 time units the pattern is highly asymmetric, indicating chaotic motion, but for the rest 100 time units, it becomes almost quasi-periodic, leading to the conclusion that the orbit has now become regular.

The above analysis reveals that as the galaxy tends to create the central nucleus, the character of most 3D regular orbits, starting near the center of the primary galaxy, changes their nature from regular to chaotic. At the same time, the velocity shows significant changes and increases as the mass of the nucleus increases. In other words, as the quiet galaxy shows this active face, the nature of motion presents major alterations. Using the above procedure, we have tested a large number of 3D orbits (approximately $10^3$) in the time-dependent model, describing the formation of a massive and dense nucleus in the core of the primary galaxy, when mass is transported from the disk. Our numerical results indicate that the character of the 3D orbits can change either from regular to chaotic and vice versa or not change at all, as the mass is transported in order to create the nucleus in the central region of the disk galaxy. In particular, from the sample of the $10^3$ tested orbits in the case of the time-dependent model, we conclude that 61\% of the orbits altered their nature from regular to chaotic, 22\% remain chaotic, 15\% remain regular, and only 2\% changed their character from chaotic to regular. Thus, it is evident that the formation of a massive nucleus leads the majority of the orbits of the 3D system to become chaotic. Whether the nature of a 3D orbit will change or not during the galactic evolution described by the set of equations (28) depends strongly on the initial conditions $\left(x_0, p_{x0}, z_0\right)$ of each orbit. Moreover, as the change of the value of the Hamiltonian (10) is negligible $\left(\Delta E_J = | E_{Jf} - E_{Ji} | \simeq 10^{-4}\right)$, we can say that the phase space is transformed to itself during the galactic evolution and, therefore, the orbits can be considered isoenergetic. In order to make this statement more clear, we present an example. If we suppose that the evolving time-dependent model was describing the 2D system, then as the change of the Hamiltonian is negligible, we could say that in the phase plane of the quiet galaxy, shown in Figure 7(b), chaotic regions would appear in the central area and it would be transformed to the phase plane shown in Figure 7(a), as the nucleus is formed in the core of the primary disk galaxy.

\section{Discussion and Conclusions}

In the present article, we have constructed a 3D gravitational model, in order to study the character of motion in a binary galactic system. In particular, we present a dynamical model which is composed of a primary disk galaxy and a small satellite companion. The motion of a test particle (star) in the gravitational field of this galactic system was studied and various techniques were used to identify regular and chaotic motion. The galaxies were assumed to be coplanar and orbiting each other in the same plane on a circular orbit. In particular, this research is indeed an extension to three dimensions of the work of Caranicolas \& Innanen (2009), who studied a similar two-dimensional binary system of interacting galaxies.

A binary system of interacting galaxies is very complex and, therefore, we need to assume some necessary simplifications and assumptions, in order to be able to study the orbital behavior of such a complicated stellar system. Thus, our model is simple and contrived, in order to give us the ability to study different aspects of the dynamical model. Nevertheless, contrived models can provide an insight into more realistic stellar systems, which unfortunately are very difficult to be studied if we take into account all the astrophysical aspects. Here, we must point out the main restrictions and limitations of our gravitational model: (1) The two galaxies (the primary and the satellite) are assumed to be coplanar, orbiting each other in the same plane on circular orbits. (2) Our dynamical model only deals with the non-dissipative components of the galaxies, stars, or possibly dark matter particles. (3) The potentials we use are rigid and do not respond to the evolving density distribution in a more realistic way. This is because our gravitational model that describes the binary system of the interacting galaxies is not self-consistent. Thus, the two galactic centers remain stationary in the rotating frame chosen. Self-consistent models are usually used when conducting $N$-body simulations. Obviously, this is out of the scope of the present paper. Once again, note that the above restrictions and limitations of our model are necessary; otherwise it would be extremely difficult, or even impossible, to apply the extensive and detailed dynamical study presented in this study. A similar gravitational model with the same limitations and assumptions was used in Zotos (2012b) in order to study the motion in a binary system of interacting galaxies, where the second galaxy is no longer treated as a point-mass satellite companion. Thus, we may conclude that the same setup has been successfully applied to describe a much more complex binary galactic system.

In our study, two different cases were investigated: the time-independent model and the time-evolving model, that is the case when mass is transported adiabatically from the disk of the galaxy to its center, forming a massive and dense nucleus. Our numerical calculation indicates that there are several factors responsible for the observed chaotic motion in the time-independent model: (i) the galactic interaction, (ii) the galactic activity, that is the presence of the nucleus, and (iii) the Lindblad resonances. Furthermore, the presence of the nucleus increases the velocities near the central region of the primary galaxy. The value of the velocity depends on the mass of the nucleus and also on the value of its scale length. Regular motion corresponds to low central velocities, while chaotic motion is characterized by high velocities. All the above observations strongly indicate that in the centers of active galaxies, chaotic motion in high velocities is expected. On the other hand, it was found that the two interacting galaxies, for large values of the distance $R$ between their centers, do not present chaotic motion, when the nucleus is not present in the core of the primary galaxy. The results of this work regarding the nature of the orbits and also the factors that affect it or change it are very similar to the corresponding outcomes obtained in Zotos (2012b) and, therefore, verify in a way our previous conclusions.

We have started our investigation from the Hamiltonian system of two degrees of freedom (2D). Our numerical calculations indicate that in this case, a large part of the phase plane is covered by chaotic orbits, while the regular regions are confined mainly near the stable retrograde periodic point. The chaotic area is larger when the primary galaxy possesses a dense and massive nucleus or when the distance $R$ between the two galaxies is small. It was also found that the velocity near the central region of the galaxy increases significantly when the nucleus is present. For lower values of the total energy and small values of the distance $R$, no chaos is observed when the nucleus is absent, while a small chaotic region appears in the presence of the nucleus. This means that low-energy stars are in chaotic orbits near the centers of active galaxies with a suitable satellite companion in a circular orbit, while in quiet galaxies they are not. Furthermore, the velocity near the center rises to high values, even for smaller values of the energy. The above discussion strongly indicates that in the centers of active galaxies, chaotic motion and high velocities are expected. This fact, combined with outcomes from previous works, shows that the majority of orbits in galaxies with dense and massive nuclei are in chaotic orbits (see Caranicolas \& Innanen 1991; Caranicolas \& Papadopoulos 2003; Zotos 2012a). This seems reasonable because theoretical results show that the nuclear force near the center increases linearly as $M_{\rm n}$ increases. Furthermore, the velocity near the galactic center strongly depends on the scale size of the nucleus $c_{\rm n}$. The smaller the $c_{\rm n}$, the higher is the velocity near the central region of an active galaxy.

To explore and understand the nature of orbits in the 3D dynamical system, we have used our knowledge obtained from the study of the 2D system. Of particular interest was the determination of the regions of initial conditions in the $\left(x, p_x, z\right)$, $p_y>0$, $\left(y=p_z=0\right)$ phase space that produce regular or chaotic 3D orbits. As the value of $p_{y0}$ was found from the Jacobi integral (10), we have used the same value of $E_J$ as in the 2D system and took initial conditions $\left(x_0, p_{x0}, z_0\right)$ such that $\left(x_0, p_{x0}\right)$ lies in the chaotic regions of the 2D system. It was found that the motion is chaotic for all permissible values of $z_0$. On the other hand, when $\left(x_0, p_{x0}\right)$ was inside a regular region around the direct and retrograde periodic points, the corresponding 3D orbits are regular for small values of $z_0$, while for larger values of $z_0$ the orbits become chaotic. The particular values of $z_0$ for which the transition from regularity to chaos in 3D orbits occurred were different for each regular region of the 2D system. Of particular interest are the results given in Table 1, where we define the average minimum value of $z_0$ near the direct and retrograde periodic points of the primary galaxy, where the transition from regularity to chaos occurs.

In this research, we have used different kinds of theoretical techniques in an attempt to explain and justify the dynamical structure of the 2D or 3D Hamiltonian system. An important role is played by the $L_z$ component of the test particle's angular momentum. It was found that the values of $\langle L_z\rangle$ for regular 3D orbits are larger than those for chaotic 3D orbits. Thus, the $L_z$ component of the angular momentum is a significant dynamical parameter connected with the regular or chaotic character of orbits in both 2D and 3D dynamical systems. In order to estimate the degree of chaos in the 2D as well as in the 3D dynamical system, we have computed the average value of the LCE for a large number of orbits with different initial conditions in the chaotic regions in each case, for a time period of $10^5$ time units. Our numerical results indicate that the degree of chaos in the 3D binary system of interacting galaxies is smaller than that in the 2D system.

It was of great interest to follow the evolution of orbits as nucleus formed in the central region of the primary disk galaxies. In this procedure, mass was transported from the disk to the nucleus and therefore the quiet galaxy became gradually active, following Equation (28). We observed that the final character of the 3D orbits strongly depends on the particular initial conditions $\left(x_0, p_{x0}, z_0\right)$. Thus, regular orbits can change to chaotic and vice versa, or maintain their character (ordered or chaotic) during the nuclear formation and since the nucleus has been formed. It was observed that a number of regular quasi-periodic orbits, starting near the central region of the primary galaxy, become chaotic. This can be seen in a total velocity versus time plot, where the asymmetric profile of the total velocity is in agreement with observational data (see Grosb{\o}l 2002), where an increase of the stellar velocity is expected, in regions with significant chaoticity. Moreover, observations show that an asymmetric velocity profile indicates chaotic motion. An interesting question is whether interactions are essential in order to trigger the mass transportation and therefore the galactic activity. This question remains unsolved as astronomers have found binary systems with disk galaxies (see Letawe et al. 2006) showing no signs of interaction at all but harboring active nuclei.

Our theoretical dynamical analysis of the binary galactic system can now be compared with observational data derived from the M51-type binary galactic systems. These systems consist of a large, primary disk galaxy and a smaller satellite companion galaxy moving in a circular orbit around their common mass center (see Klimanov \& Reshetnikov 2001). We apply our results to the binary system composed of the primary galaxy NGC 5829 and its small satellite IC 4526. Figure 29 depicts a real image of this binary system. The primary galaxy has an absolute magnitude of $M_V=-20.66$. In order to estimate the disk's radius, we shall use the empirical relation
\begin{equation}
M_V = -6 \log A + 7.14,
\end{equation}
which connects the absolute magnitude to the projected major axis of the primary galaxy $A$ in pc (see Bowers \& Deeming 1984). From relation (29), we obtain the value $A$ = 43 kpc. If we adopt the values of the parameters given in Figures 7(a--b), then one should expect a large number of the orbits to be chaotic in this binary system of interacting galaxies. We can also observe that the orbit of the satellite is close to the boundary of the galactic disk. This result is in agreement with the observational data as well.
\begin{figure}[!tH]
\centering
\resizebox{\hsize}{!}{\rotatebox{0}{\includegraphics*{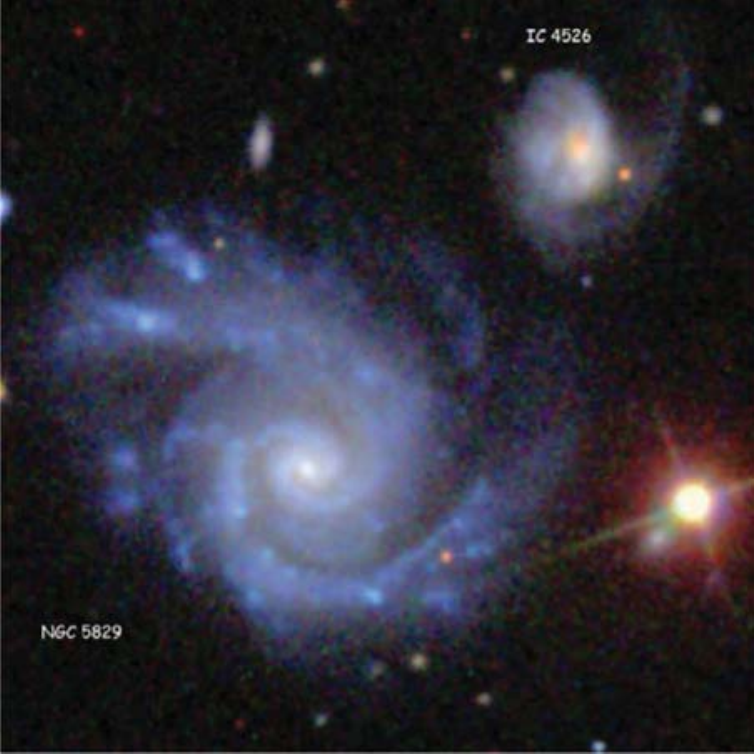}}}
\caption{\small A real image of the binary galactic system composed of the primary galaxy NGC 5829 and its small satellite companion IC 4526.}
\end{figure}

When studying galactic dynamics, it is invariably found that the stellar rotation velocity remains constant, or flat, with increasing distance away from the galactic center (see Cantinella, Giovanelli, \& Haynes 2006). This result is highly counterintuitive since, based on Newton's law of gravity, the rotational velocity would steadily decrease for stars farther away from the galactic center. By this particular argument, the flat rotational curves seem to imply that each galaxy must be surrounded by significant amounts of dark matter. It has been postulated and generally accepted that the dark matter would have to be located in a massive halo enshrouding each galaxy (see Caranicolas \& Zotos 2009b, 2011a, 2011b). The first real surprise in the study of dark matter lay in the outermost parts of galaxies known as galaxy halos. Here, there is negligible luminosity, yet there are occasional orbiting gas clouds, which allow us to measure rotation velocities and distances. The rotation velocity was found not to decrease with increasing distance from the galactic center, implying that the mass distribution of the galaxy cannot be considered like the light distribution. The mass must continue to increase since the rotation velocity satisfies
\begin{equation}
\Theta ^2 = \frac{G M}{r},
\end{equation}
where $M$ is the mass within radius $r$; we infer that $M$ increases proportionally to $r$. This rise appears to stop at about 50 kpc, where halos appear to be truncated. We infer that the mass--luminosity ratio of the galaxy $(M/L)$, including its disk halo, is about 5 times larger than estimated for the luminous inner region or equal to about 50. The rotation curve shown in Figure 3, corresponding to our disk galaxy, seems to maintain high values of the angular velocity $\Theta$ for large distances from the galactic center. Thus, we may say that our dynamical model can interpret the presence of dark matter.

Forty years ago, galactic activity and interactions between galaxies were viewed as unusual and rare. Nowadays, they seem to be segments in the life of many galaxies. From the astrophysical point of view, in the present work, we have tried to connect galactic activity and galactic interactions with the nature of orbits (regular or chaotic) and also with the behavior of the velocities of stars in the primary galaxy. We consider the outcomes of the present research to be an initial effort, in order to explore the dynamical structure of the 3D binary stellar system in more detail. As results are positive, further investigation will be initiated to study all the available phase spaces, including orbital eccentricity (elliptic orbits) of the small companion galaxy and its inclinations to the primary galaxy. Moreover, we shall try to use the outcomes obtained from this initial and simple dynamical model to conduct computer $N$-body simulations in a binary system of interacting galaxies (a primary galaxy with a satellite companion), in order to reveal changes in their orbital properties through merger processes and tidal effects, which are obviously out of the scope of this study.

\section*{Acknowledgments}

The author would like to express his warmest thanks to the anonymous referee for careful reading of the manuscript and for his very useful and illuminating suggestions and comments that greatly improved the quality and also the clarity of the present article.

\end{document}